\documentclass[prd,onecolumn,amsmath,showpacs,nofootinbib,11pt]{revtex4}
\usepackage{graphicx}
\usepackage{float}
\usepackage{dcolumn}
\usepackage{bm}
\begin{document}
\newcommand{\hs}{\hspace*{0.5cm}}
\newcommand{\vs}{\vspace*{0.5cm}}
\newcommand{\be}{\begin{equation}}
\newcommand{\ee}{\end{equation}}
\newcommand{\bea}{\begin{eqnarray}}
\newcommand{\eea}{\end{eqnarray}}
\newcommand{\ben}{\begin{enumerate}}
\newcommand{\een}{\end{enumerate}}
\newcommand{\bde}{\begin{widetext}}
\newcommand{\ede}{\end{widetext}}
\newcommand{\nn}{\nonumber}
\newcommand{\crn}{\nonumber \\}
\newcommand{\Tr}{\mathrm{Tr}}
\newcommand{\non}{\nonumber}
\newcommand{\noi}{\noindent}
\newcommand{\al}{\alpha}
\newcommand{\la}{\lambda}
\newcommand{\bet}{\beta}
\newcommand{\ga}{\gamma}
\newcommand{\va}{\varphi}
\newcommand{\om}{\omega}
\newcommand{\pa}{\partial}
\newcommand{\+}{\dagger}
\newcommand{\fr}{\frac}
\newcommand{\bc}{\begin{center}}
\newcommand{\ec}{\end{center}}
\newcommand{\Ga}{\Gamma}
\newcommand{\de}{\delta}
\newcommand{\De}{\Delta}
\newcommand{\ep}{\epsilon}
\newcommand{\varep}{\varepsilon}
\newcommand{\ka}{\kappa}
\newcommand{\La}{\Lambda}
\newcommand{\si}{\sigma}
\newcommand{\Si}{\Sigma}
\newcommand{\ta}{\tau}
\newcommand{\up}{\upsilon}
\newcommand{\Up}{\Upsilon}
\newcommand{\ze}{\zeta}
\newcommand{\ps}{\psi}
\newcommand{\Ps}{\Psi}
\newcommand{\ph}{\phi}
\newcommand{\vph}{\varphi}
\newcommand{\Ph}{\Phi}
\newcommand{\Om}{\Omega}

\title{Phenomenology of the $SU(3)_C \otimes SU(2)_L \otimes SU(3)_R \otimes U(1)_X$ gauge model}

\author{P. V. Dong}\email{pvdong@iop.vast.vn}
\address{Institute of Physics, Vietnam Academy of Science and Technology, 10 Dao Tan, Ba
Dinh, Hanoi, Vietnam}

\author{D. T. Huong}\email{dthuong@iop.vast.vn}
\address{Institute of Physics, Vietnam Academy of Science and Technology, 10 Dao Tan, Ba
Dinh, Hanoi, Vietnam}

\author{D. V. Loi}\email{dvloi@grad.iop.vast.vn}
\address{Faculty of Mathematics-Physics-Informatics, Tay Bac University,
Quyet Tam, Son La, Vietnam}

\author{N. T. Nhuan}\email{ntnhuan@grad.iop.vast.vn}
\address{Graduate University of Science and Technology, Vietnam Academy of Science and Technology, 18 Hoang Quoc Viet, Cau Giay, Hanoi, Vietnam}

\author{N. T. K. Ngan}
\email{ntkngan@ctu.edu.vn}\affiliation{Department of Physics, Cantho University, 3/2 Street, Ninh Kieu, Cantho, Vietnam}

\begin{abstract}
We study the left-right asymmetric model based on $SU(3)_C\otimes
SU(2)_L \otimes SU(3)_R\otimes U(1)_X $ gauge group, which improves the theoretical and phenomenological aspects of the known left-right symmetric model. This new gauge symmetry yields that the fermion generation number is three, and the tree-level flavor-changing neutral currents arise in both gauge and scalar sectors. Also, it can provide the observed neutrino masses as well as dark matter automatically. Further, we investigate the mass spectrum of the gauge and scalar fields. All the gauge interactions of the fermions and scalars are derived. We examine the tree-level contributions of the new neutral vector, $Z'_{R}$, and new neutral scalar, $H_2$, to flavor-violating neutral meson mixings, say $K$-$\bar{K}$, $B_d$-$\bar{B}_d$, and $B_s$-$\bar{B}_s$, which strongly constrain the new physics scale as well as the elements of the right-handed quark mixing matrices. The bounds for the new physics scale are in agreement with those coming from the $\rho$-parameter as well as the mixing parameters between $W,\ Z$ bosons and new gauge bosons.   
\end{abstract}

\date{\today}

\maketitle

\section{\label{intro}Introduction}

In the standard model, the neutral currents of $\gamma$ and $Z$ conserve every flavor at the tree-level, whereas the charged current of $W$ changes quark flavors through the CKM matrix (where lepton flavors are separately conserved). This directly leads to quark-flavor violating processes such as neutral meson mixings, $K$-$\bar{K}$, $D$-$\bar{D}$, $B_{d}$-$\bar{B}_{d}$, and $B_{s}$-$\bar{B}_{s}$, and rare meson decays, $B_s \rightarrow \mu^+ \mu^-$, $B_s \rightarrow \varphi \mu^+ \mu^-$, $B_d \rightarrow K(K^*) \mu^+ \mu^-$, and others. All such standard model predictions have been experimentally tested so far, and that they are globally compatible with the existing data~\cite{data}. However, with the reduced experimental errors as well as enhanced QCD and EW precision computations, a number of tensions have recently been found at 2--3$\sigma$ levels corresponding to individual processes \cite{bsmumu,lhcb1,lhcb2,lhcb3,lhcbcombi}. Whilst some of them might be due to statistical fluctuations/errors, it does not exclude a possibility that they reveal some new physics. Further, the standard model cannot explain the small, nonzero neutrino masses and lepton-flavor mixings. It also fails to address dark matter that occupies roundly 25\% of the mass-energy density of the universe.      

The minimal left-right symmetric model based on $SU(3)_C\otimes SU(2)_L  \otimes SU(2)_R \otimes U(1)_{B-L}$ gauge group
is one of the most attractive extensions of the standard model \cite{LR1,NLR}. A motivation of the model
is that the parity is exact but its asymmetry as seen in the weak interaction is due to the spontaneous breaking of $SU(2)_R$ at some large energy scale. It also plays an important role in developing the theories  
of neutrino masses, well-known as seesaw mechanisms, and that non-zero neutrino masses were suggested long before the experimental confirmations. Particularly, the phenomenological consequences of the new particles that contribute to the meson mixing systems as well as rare meson decays were studied in \cite{PheLR}. The contribution of the right current that addresses $V_{ub}$ problem was also discussed in \cite{vubpro}. Generally, the experimental bounds would require the left-right scale to be in the TeV region, and the explicit left-right asymmetries should be imposed in order to fit most, but not all, the data, which may be well tested at the LHC run II. As the standard model, the minimal left-right symmetric model cannot solve the dark matter issue.    

Furthermore, the minimal left-right symmetric model did not contain the necessary ingredients for solving the 750 GeV diphoton excess naturally \cite{exdiphoton}. Although the resonance was subsequently proved as statistic fluctuations \cite{noexdiphoton}, the guidance for going beyond this model to address new physics anomalies like that is still worth studying. Literately, the proposals \cite{diphoton} that enlarged only the particle content are not considered here since they included the new fields by hand, and obviously it is not naturally on both phenomenological and theoretical grounds. However, the proposals \cite{3L3R} that extended the gauge group can show alternative important results since it manifestly follows a gauge principle. Indeed, it was shown that the diphoton anomaly might be associated with fundamental left-right asymmetries, and thus the three theories of which were proposed, corresponding to the gauge symmetries, $SU(3)_C\otimes SU(M)_L\otimes SU(N)_R\otimes U(1)_X$ (3-$M$-$N$-1), for $(M,N)=(2,3),\ (3,2)$, and $(3,3)$, respectively. Here, the left-right asymmetry is either explicitly recognized for $M\neq N$ or spontaneously produced after the gauge symmetry breaking for $M=N$. The diphoton excess was the new scalar fields, produced/decayed as mediated by the new fermions, which all transform as fundamental components, in the quotient space $[SU(M)_L\otimes SU(N)_R]/[SU(2)_L\otimes SU(2)_R]$, enlarged from those of the minimal left-right symmetric model. 

However, the new physics scales were generally low, below a few TeVs, and the characteristic electric charge parameter was rarely big, in order to explain the large diphoton signal strength. Because the massive diphoton signals were absent, the new physics scales must be large, above those bounds, which are also needed to evade other constraints discussed hereafter (as also noted in \cite{3L3R}), and the electric charge parameter is not necessarily large beyond the usual electric charges. Simultaneously, as shown in this work, the fundamental left-right asymmetries as proposed provide automatically the tree-level flavor-changing neutral currents (FCNCs) through the gauge and Yukawa interactions, which may be the new source for addressing the B physics anomalies and others, which dominate over those loop-induced by the minimal left-right symmetric model. Additionally, the new gauge symmetries that reflect the left-right asymmetries can supply dark matter naturally by the mean that dark matter candidates and their stability mechanism and relic density automatically arise from the gauge principles.       
  
In this work, we take the most simple theory among the three mentioned ones into account, which is given by the $SU(3)_C\otimes SU(2)_L  \otimes SU(3)_R \otimes U(1)_{X}$ (3-2-3-1) gauge symmetry. Note that the two others, namely the 3-3-2-1 and 3-3-3-1 models, include an extension, $SU(2)_L\rightarrow SU(3)_L$, besides the corresponding enlargements of the weak hypercharge. Therefore, we see that the left-handed fermion-content and symmetry are the same standard model and minimal left-right symmetric model, but the right sector is extended, explicitly violating a symmetry between the left and right, the so-called left-right asymmetry. This approach predicts the three fermion generations as observed as a result of $SU(3)_R$ anomaly cancelation and QCD asymptotic freedom. The new FCNCs come from two distinct sources as loop induced by $W_R$ and a charged Higgs boson and tree-level contributed by non-universal couplings of $Z'_R$ and a neutral Higgs boson ($H_2$) with ordinary quarks. The former is similar to the minimal left-right symmetric model, which is negligible as suppressed by loop factors. Whereas, the latter may dominate which is in charge to interpret the mentioned flavor-physics anomalies. Additionally, the new gauge symmetry, $SU(3)_R\otimes U(1)_X$, may define a nontrivial W-parity as well as W-odd matter content responsible for dark matter, which is quite similar to the 3-3-1-1 model \cite{d3311}.    
 
Let us stress that the interesting feature of the considering model is that the right-handed quarks of the first and second generations transform under the gauge symmetry differently from the third generation, which directly leads to the tree-level FCNCs caused by the only right-handed quarks when coupling to the $H_2$ scalar and $Z'_R$ gauge boson. This property does not exist in the 3-3-1 models as well as the other left-right theories. The former including the two remaining left-right asymmetric theories have a similar property but caused by the left-handed quarks \cite{331,3L3R}. Unlike those theories, the model under consideration yields that the relevant observables depend only on the new energy scale and the right-handed quark mixing matrices, $V_{uR}$ and $V_{dR}$, which are not constrained by the standard model (i.e., they act as arbitrary parameters). Further, this proposal implies the neutrino masses via seesaw mechanisms like the minimal left-right symmetric model. The electric charge operator is directly related to the baryon-minus-lepton charge ($B-L$), which is unlike the 3-3-1-1 model \cite{d3311}. The characteristic electric charge parameter of the model also defines $B-L$ charge for the new particles, by which a class of wrong $B-L$ particles are naturally recognized, which transform nontrivially under a residual discrete gauge symmetry, called W-parity. The new gauge and Higgs bosons might also be the subjects for the dijet, Drell-Yan, and diboson searches by the LHC experiments. For the whole purposes, we will identify the scalars and gauge bosons as well as calculating all the necessary gauge interactions.

The rest of this paper is organized as follows: In Sec. \ref{model}, we give a detailed review of the model with stressing on dark matter and FCNCs. Sections \ref{higgssec} and \ref{gaugesec} study the mass spectrum of the scalar and gauge boson fields, respectively. The gauge interactions of fermions and scalars are considered in Sec. \ref{interactionsec}.  Section \ref{flavorsec} is devoted to the FCNCs which are directly mediated by the new neutral gauge and scalar
fields. The mixing effects in the gauge and scalar sectors are also discussed therein. Finally, we summarize our results and conclude this work in Sec. \ref{conclusion}. 

\section{\label{model} The model}

As mentioned, the gauge symmetry of the model is defined by $SU(3)_C \otimes SU(2)_L \otimes SU(3)_R \otimes
 U(1)_X$, where the first group factor is the ordinary QCD symmetry, while the last three are an extension of
the electroweak symmetry, which contains that of the minimal
 left-right symmetric model as a subgroup. However, the considering model does not conserve a left-right symmetry, $Z_2$, that interchanges the left and right gauge groups as well as their corresponding field contents, i.e. it presents an explicit left-right asymmetry. 
 
 The electric charge operator is embedded in the gauge symmetry as follows \cite{3L3R}
\bea Q =T_{3L}+ T_{3R}+ \beta T_{8R}+X, \label{electric1} \eea where $T_{aL}$ ($a=1,2,3$), $T_{iR}$
($i=1,2,3,...,8$),  and $X$ are $SU(2)_L$, $SU(3)_R$, and $U(1)_X$ generators, respectively. $\beta$ can be expressed via an electric charge parameter ($q$) as $\beta=-(2q+1)/\sqrt{3}$. The baryon-minus-lepton charge is embedded as $\fr 1 2 (B-L)=\beta T_{8R}+X$. Hence, depending on the embedding parameter $\beta$ (or $q$), this model may automatically provide dark matter candidates, which are stabilized by a W-parity, \be P=(-1)^{3(B-L)+2s}=(-1)^{6(\beta T_{8R}+X)+2s},\ee as residual gauge symmetry, similarly to the 3-3-1-1 model \cite{d3311} (see below).     
 
 The fermion content that is anomaly free is given by \cite{3L3R}
 \bea
\psi_{aL}=\left(%
\begin{array}{c}
  \nu_{aL} \\
  e_{aL}\\
\end{array}%
\right)\sim \left(1,2,1,-\frac{1}{2}\right),\hs \hs \psi_{aR}=\left(%
\begin{array}{c}
  \nu_{aR} \\
 e_{aR} \\
  E_{aR}^q \\
\end{array}%
\right)\sim \left(1,1,3,\frac{q-1}{3}\right),
 \eea
 \bea
Q_{3L}= \left(%
\begin{array}{c}
  u_{3L} \\
  d_{3L} \\
\end{array}%
\right) \sim \left(3,2,1,\frac{1}{6}\right),\hs \hs Q_{3R}= \left(%
\begin{array}{c}
  u_{3R} \\
  d_{3R} \\
  J^{q+\frac{2}{3}}_{3R} \\
\end{array}%
\right)\sim \left(3,1,3,\frac{q+1}{3}\right),
 \eea
 \bea
 Q_{\al L} =\left(%
\begin{array}{c}
  u_{\al L} \\
  d_{\al L} \\
\end{array}%
\right)\sim \left(3,2,1,\frac{1}{6}\right),\hs \hs Q_{\al R}=\left(%
\begin{array}{c}
  d_{\al R} \\
  -u_{\al R} \\
  J_{\al R}^{-q-\frac{1}{3}} \\
\end{array}%
\right)\sim \left(3,1,3^*,-\frac{q}{3}\right),
 \eea
 \bea
 E^q_{aL}\sim (1,1,1,q), \hs J^{q+\fr 2 3}_{3L}\sim \left(3,1,1,q+\fr{2}{3}\right), \hs J^{-q-\fr 1 3}_{\al L} \sim
 \left(3,1,1,-q-\fr{1}{3}\right),
 \eea
where $a=1,2,3$ and $\al=1,2$ are generation indices. The numbers in the parentheses denote the quantum numbers up on the 3-2-3-1 subgroups, respectively.    

We see that the proposal of $SU(3)_R$ leads to not only the existence of the right-handed neutrinos which induce the neutrino masses via seesaw mechanisms, but also the new leptons $E_a$ and exotic quarks $J_a$ which might yield interesting phenomena. Indeed, note that $E_a$ and $J_a$ have $B-L$ charge equal to two times their electric charges, i.e. $[B-L](E_a)=2q$, $[B-L](J_3)=2(q+2/3)$, and $[B-L](J_\al)=2(-q-1/3)$. Therefore, the model recognizes a nontrivial W-parity for wrong $B-L$ particles that include $E,J$ and others, called W-particles, responsible for dark matter if $q\neq (2m-1)/6=\pm1/6,\pm1/2,\pm 5/6,\pm 7/6,\cdots$ for $m$ integer \cite{d3311}. Here, W-particles have $P=P^+$ or $P^-$, with $P^\pm\equiv (-1)^{\pm(6q+1)}\neq 1$, while the remaining particles that include the standard model particles and some new ones have $P=1$, called normal particles. Particularly, the model with ordinary charge $q=m/3=0,\pm1/3,\pm2/3,\pm1,\cdots$ belongs to this class, which yields W-parity as R-parity and W-particles as R-odd particles (the simplest, but realistic, version is if $q=0$).        

Provided that the right-handed fermions are arranged in the fundamental representations of $SU(3)_R$, the $SU(3)_R$ anomaly cancelation demands the number of triplets equaling that of antitriplets. Thus, the generation number must be a multiple of three. Since the extra quarks are included to complete the representations, the QCD asymptotic freedom requires the generation number to be less than or equal to five. Hence, the generation number is three, as expected. Furthermore, the right-handed quarks of the third generation transform differently from those of the first two generations. This leads to the tree-level quark FCNCs due to the interactions with the new gauge bosons of $T_{8R}$ and $X$ as well as new neutral scalars (shown below). This demonstrates that the new dominant FCNCs all arise from such an explicit left-right symmetry violation.     

To break the gauge symmetry and generate appropriate masses for the particles, the scalar multiplets are introduced as
 \bea
S &=& \left(%
\begin{array}{ccc}
  S_{11}^0 & S_{12}^+ & S_{13}^{-q} \\
  S_{21}^- & S_{22}^0 & S_{23}^{-q-1} \\
\end{array}%
\right) \sim \left(1,2,3^*,-\frac{2q+1}{6}\right),\\ 
 \phi &=& \left(%
\begin{array}{c}
  \phi_1^{-q} \\
  \phi_2^{-q-1} \\
  \phi_3^0 \\
\end{array}%
\right)\sim \left(1,1,3,-\frac{2q+1}{3}\right), \\  
\Xi &=& \left(%
\begin{array}{ccc}
  \Xi^0_{11} &\frac{ \Xi_{12}^-}{\sqrt{2}} &\frac{ \Xi_{13}^q }{\sqrt{2}}\\
  \frac{ \Xi_{12}^-}{\sqrt{2}} & \Xi_{22}^{--} & \frac{\Xi_{23}^{q-1}}{\sqrt{2}} \\
  \frac{\Xi_{13}^q}{\sqrt{2}} & \frac{\Xi_{23}^{q-1}}{\sqrt{2}} & \Xi_{33}^{2q}\\
\end{array}%
\right)\sim \left(1,1,6,\frac{2(q-1)}{3}\right),
 \eea
which have the corresponding vacuum expectation values (VEVs), \bea \langle S\rangle  &=&\frac{1}{\sqrt{2}}\left(%
\begin{array}{ccc}
  u & 0 & 0 \\
  0 & v & 0 \\
\end{array}%
\right), \hs  \langle \phi\rangle =\frac{1}{\sqrt{2}}\left(%
\begin{array}{c}
  0 \\
  0 \\
 w\\
\end{array}%
\right), \hs  \langle \Xi\rangle =\frac{1}{\sqrt{2}}\left(%
\begin{array}{ccc}
  \Lambda & 0 & 0 \\
  0 & 0 & 0 \\
  0 & 0 & 0 \\
\end{array}%
\right).\eea
As mentioned in \cite{3L3R}, if one introduces a scalar triplet $\Delta$ the neutrinos get masses through a combination of the type I and II seesaw mechanisms; otherwise, the only type I seesaw mechanism is presented. Since both cases can fit the data, we would not include $\Delta$ for simplicity. The W-fields include $\phi_{1,2}$, $S_{13,23}$, and $\Xi_{13,23}$. The other scalars are normal fields.  

The gauge symmetry is broken via two steps, 
\bc
\begin{tabular}{c}  
$SU(3)_C\otimes SU(2)_L\otimes SU(3)_R \otimes U(1)_X$\\
$  \downarrow w,\Lambda$\\
$SU(3)_C\otimes SU(2)_L \otimes U(1)_Y\otimes P$\\
$\downarrow u, v$\\ 
$SU(3)_C\otimes U(1)_Q\otimes P$
\end{tabular}
\ec
Here, the VEV of $\phi$ ($w$) provides the masses for new leptons and exotic quarks, while the VEV of $\Xi$ ($\La$) provides the Majorana masses for right-handed neutrinos. Both the VEVs $w,\La$ give the masses for new gauge bosons. The VEVs of $S$ ($u$, $v$) generate the masses for ordinary charged-leptons, quarks, and weak gauge bosons as well as Dirac masses for neutrinos. Subsequently, the small neutrino masses are induced via the seesaw mechanism as mentioned. Additionally, after the first step of symmetry breaking, the W-parity is defined along with the standard model symmetry due to the VEV $\La$ \cite{d3311}. Note that $w,u,v$ do not break $B-L$, whereas $\La$ breaks this charge which defines the Majorana masses and W-parity. Thus, the observed neutrino masses and dark matter stability are strongly correlated, as originating from the $B-L$ gauge symmetry breaking. To be consistent with the standard model, we must impose $u,v \ll w,\La$.   

The total Lagrangian takes the form, 
\bea
\mathcal{L}= \mathcal{L}_{\mathrm{kinetic}} + \mathcal{L}_{\mathrm{Yukawa}}-V_{\mathrm{scalar}},
\eea
where the first part includes kinetic terms and gauge interactions, which will be obtained latter. The second and last parts correspond to the Yukawa Lagrangian and scalar potential, respectively, which are obtained by 
 \bea
 \mathcal{L}_{{\mathrm{Yukawa}}} & =&  h^l_{ab} \bar{\psi}_{aL}S \psi_{bR}
 +h^{R}_{ab}
\bar{\psi}^c_{aR} \Xi^\dagger \psi_{bR} + h_{a3}^q\bar{Q}_{aL}S Q_{3R}+h^q_{a \beta}\bar{\tilde{Q}}_{a L} S^* Q_{\beta R}
\crn
&& + h_{ab}^E\bar{E}_{aL}\phi^\dagger \psi_{bR} + h^J_{33}\bar{J}_{3L}\phi^\dagger Q_{3R} + h^J_{\al \beta} \bar{J}_{\al L} \phi^T Q_{\beta R}+H.c., \label{231}\\ V_{\mathrm{scalar}} &= &\mu_S^2 \Tr(S^\dag S)+\la_{1S}[\Tr (S^\dag
S)]^2+\la_{2S}\Tr(S^\dag SS^\dag S) +\mu_\Xi^2\Tr(\Xi^\dag \Xi)\crn 
&& +\la_{1\Xi}[ \Tr(\Xi^\dag \Xi)]^2 +\la_{2 \Xi}\Tr( \Xi^\dag \Xi \Xi^\dag \Xi)+\mu_\phi^2 \phi^\dag \phi +\la_\phi (\phi^\dag \phi)^2\crn
&& +\la_1(\phi^\dag S^\dag S\phi)+\la_2\Tr(S^\dag S \Xi \Xi^\dag)+\la_3(\phi^\dag\Xi\Xi^\dag\phi)+\la_4
(\phi^\dag \phi) \Tr(S^\dag S)\crn
&& +\la_5 (\phi^\dag \phi) \Tr(\Xi^\dag \Xi)+ \la_6\Tr(\Xi^\dag \Xi) \Tr(S^\dag
S) +  (f S\phi^*S+ H.c.),  \label{232} \eea where ``$\Tr$'' is the trace operator. Here, note that $\tilde{Q}_L\equiv i\sigma_2 Q_L$ transforms as $2^*$ under $SU(2)_L$, i.e. $\tilde{Q}_{L}\rightarrow U^*_L \tilde{Q}_L$. Also, we have $S\rightarrow U_LS U^\dagger_R$, $\Xi\rightarrow U_R \Xi U^T_R$, and $Q_{\al R}\rightarrow U^*_R Q_{\al R}$, under $SU(2)_L\otimes SU(3)_R$. Observe that the third generation of quarks interacts with the scalars with the forms differently from those for the first two quark generations, which does not happen with leptons as well as not analogous to the case of the minimal left-right symmetric model. As referred to \cite{3L3R} the potential parameters have been redefined for easily reading, and the $f$, $\la_{1,2,3}$ couplings have been imposed for generalization, which were skipped in the previous study.         

The gauge sector contains two W-fields as $X^{\pm q}_R$ and $Y^{\pm(q+1)}_R$ as coupled to $\fr{1}{\sqrt{2}}(T_{4R}\mp i T_{5R})$ and $\fr{1}{\sqrt{2}}(T_{6R}\mp i T_{7R})$, respectively. The other gauge bosons are normal fields. Summarizing all the W-fields, we see that the model provides dark matter candidates if $q=0,\pm 1$ (note that the candidate must be electrically neutral). The model with $q=0$, the candidates are $E^0$ or $X^0_R$ or some combination of ($\phi^0_1$, $S^0_{13}$, $\Xi^0_{13}$). The model with $q=-1$, the candidates are $Y^0_R$ or some combination of $(\phi^0_2,S^0_{23})$. The model with $q=1$, the candidates are only $\Xi^0_{23}$. The dark matter candidate must be the lightest W-particle, called LWP, which is stabilized by W-parity. {\it Prove}: one has an interaction of $r$ $P^+$-fields and $s$ $P^-$-fields, where $r$ and $s$ are integer. Since $P$ is conserved, it follows $(P^+)^r(P^-)^s=1$, which is valid only if $r=s$. In other words, $P^+$ and $P^-$ always appear in pairs in interactions, which is analogous to superpartices in supersymmetry. 

A detailed study of the three mentioned versions for dark matter phenomenologies is out of the scope of this work, which should be published elsewhere \cite{dhip}. In the followings, we identify physical particles, calculate all interactions, and present selected phenomena for the general model.         

\section{\label{higgssec}Scalar sector}

Let us expand the neutral scalar fields ($S_{11}^0, S_{22}^0, \phi_3^0, \Xi_{11}^0$) around the mentioned VEVs as  
\bea
S &=&  \left(\begin{array}{ccc}  \frac{u+S_1+iA_1}{\sqrt{2}}& S_{12}^+ & S_{13}^{-q} \\
S_{21}^- &  \frac{v+S_2+iA_2}{\sqrt{2}} & S_{23}^{-q-1} \\
\end{array}\right),\label{kthiggs} \\  
\phi &=& \left(\begin{array}{c}
  \phi_1^{-q} \\
  \phi_2^{-q-1} \\
  \frac{w+ S_3+iA_3}{\sqrt{2}} \\
\end{array}\right), \hs   
\Xi = \left(\begin{array}{ccc}
\frac{\Lambda+S_4+iA_4}{\sqrt{2}} &\frac{ \Xi_{12}^-}{\sqrt{2}} &\frac{ \Xi_{13}^q }{\sqrt{2}}\\
\frac{ \Xi_{12}^-}{\sqrt{2}} & \Xi_{22}^{--} & \frac{\Xi_{23}^{q-1}}{\sqrt{2}} \\
\frac{\Xi_{13}^q}{\sqrt{2}} & \frac{\Xi_{23}^{q-1}}{\sqrt{2}} & \Xi_{33}^{2q}\\
\end{array}\right). \label{higgs} 
\eea
To find the potential minimization and scalar mass spectrum, we correspondingly expand the original potential terms up to the second order terms of the component fields given above and then sum all the resulting terms that have the same order in fields. Therefore, the scalar potential is divided into $V(S, \phi, \Xi) = V_{\mathrm{min}} + V_{\mathrm{linear}} + V_{\mathrm{mass}} + V_{\mathrm{interaction}}$, where all the interactions are grouped to $V_{\mathrm{interaction}}$, which need not to determine. $V_{\mathrm{min}}$ is the minimum of the potential, which is independent of the fields as well as only contributing to the vacuum energy.
$V_{\mathrm{linear}}$ contains all the terms that depend linearly on the fields, and that the gauge invariance requires $V_{\mathrm{linear}}=0$ which leads to the minimization conditions as follows
\bea
\mu_S^2 u+(\lambda_{1S}+\lambda_{2S})u^3-\sqrt{2}fvw+\fr 1 2 u[2 \lambda_{1S}v^2+\lambda_{4} w^2+(\lambda_2+\lambda_{6})\Lambda^2]&=&0,\label{bsd1} \\
\mu_S^2 v+(\lambda_{1S}+\lambda_{2S})v^3-\sqrt{2}fuw+\fr 1 2 v(2 \lambda_{1S}u^2+\lambda_{4} w^2+\lambda_{6}\Lambda^2)&=&0,\label{bsd2}\\
\mu_\phi^2 w+\lambda_\phi w^3-\sqrt{2}fuv+\fr 1 2 w[\lambda_{4} (u^2+v^2)+\lambda_{5}\Lambda^2]&=&0,\\
\mu_\Xi^2+(\lambda_{1\Xi}+\lambda_{2\Xi})\Lambda^2+\fr 1 2 [(\lambda_2+\lambda_{6})u^2+\lambda_{\Xi S}v^2+\lambda_{5}w^2]&=&0.
\label{minimal}\eea
$V_{\mathrm{mass}}$ consists of the terms that depend quadratically on the fields, given in the form, $V_{\mathrm{mass}} =V^{S}_{\mathrm{mass}}+ V_{\mathrm{mass}}^{A}+ V^{\mathrm{charged}}_{\mathrm{mass}}$, where the first two terms describe the CP even and CP odd scalar fields respectively, while the last one contains the charged scalar fields. 

Substituting the minimization conditions into the scalar potential, $V^{S}_{\mathrm{mass}}$ is given by 
\bea
V^{S}_{\mathrm{mass}} = \frac{1}{2}\left(\begin{array}{cccc}
S_1 &
S_2&
S_3 & S_4
\end{array}\right) M_S^2 \left(\begin{array}{cccc}
S_1 &
S_2&
S_3 & S_4
\end{array}\right)^T,
\eea
where $M_S^2$ is  
\bea
 \left(\begin{array}{cccc}
2(\lambda_{1S}+\lambda_{2S})u^2+\frac{\sqrt{2}fvw}{u} & 2\lambda_{1S}uv-\sqrt{2}fw & -\sqrt{2}fv+\lambda_4 uw & (\lambda_2+\lambda_6)u\Lambda \\
2\lambda_{1S}uv-\sqrt{2}fw&2 (\la_{1S}+ \la_{2S})v^2-\la_{2S} u^2+\fr{\la_2u^2 \La^2}{2(v^2-u^2)}&-\sqrt{2}fu+\la_4 vw&\la_6 v \La \\
-\sqrt{2}fv+\la_4 uw & -\sqrt{2}fu+\la_4 vw&\frac{\sqrt{2} f uv}{w}+2 \la_\phi w^2&\la_5 w \La \\
(\lambda_2+\lambda_6)u\Lambda & \la_6 v \La&\la_5 w \La&2(\la_{1\Xi}+\la_{2\Xi})\La^2
\end{array}\right). \nonumber  
\eea
Note that $f$ is a mass parameter satisfying 
\bea
f= -\frac{\la_{2S } uv}{\sqrt{2} w} -\frac{\la_2 u v \La^2}{2 \sqrt{2}(u^2-v^2) w},
\label{tuyentinh1}\eea which is derived from the minimization conditions (\ref{bsd1}) and (\ref{bsd2}).  
Because of $ u,v \ll w, \La$, the parameter $f$ is large in the $w,  \La$ scale.
At the leading order, $u,v\ll w,\La,f$, the above mass matrix implies a massless scalar field,
$ H_1 = \fr{uS_1 + vS_2}{\sqrt{u^2+v^2}}$, and three heavy scalar fields with masses given by  
 \bea H_2 & =&  \fr{-vS_1 + uS_2}{\sqrt{u^2+v^2}}, \hs  m^2_{H_2} = \fr{\lambda_2(u^2+v^2)\Lambda^2}{2(v^2-u^2)}, \crn \nonumber
H_3 &=& c_\varphi S_3 - s_\varphi S_4, \hs  m^2_{H_3}= \la_\phi w^2+(\lambda_{1\Xi}+\lambda_{2\Xi})\La^2 - \sqrt{ [(\lambda_{1\Xi}+\lambda_{2\Xi})\La^2-\la_\phi w^2 ]^2+\lambda_{5}^2 w^2\Lambda^2},\crn
 H_4 &=& s_\varphi S_3 + c_\varphi S_4, \hs
 m^2_{H_4}= \la_\phi w^2+(\lambda_{1\Xi}+\lambda_{2\Xi})\La^2 + \sqrt{[(\lambda_{1\Xi}+\lambda_{2\Xi})\La^2-\la_\phi w^2 ]^2+\lambda_{5}^2 w^2\Lambda^2}, \nonumber    \eea
where we have denoted $c_\varphi = \cos \varphi$, $s_\varphi= \sin \varphi$, and so forth, with \be t_{2\varphi} = \fr{\lambda_{5}w\Lambda}{(\lambda_{1\Xi}+\lambda_{2\Xi}) \La^2-\la_\phi w^2}.      \ee
At the next-to-leading order, the Higgs masses, $ m^2_{H_i}$ ($i =1,2,3,4$), are contributed by the $u^2, v^2$ terms. Particularly, the $H_1$ mass is approximated as 
\bea
 m_{H_1}^{ 2} =2(\lambda_{1S}+\lambda_{2S})u^2-\lambda_{2S}v^2.
\eea
It is easily realized that the light state, $H_1$, is identical to the standard model Higgs boson. Whereas, the heavy states, $H_{2,3,4}$, are new particles with the masses as given in $w,\La $ scales. 

The mass terms of the pseudoscalars, $A_1, A_2, A_3, A_4$, are given by 
\bea
 V_{\mathrm{mass}}^{A}= \fr 1 2 \left(\begin{array}{cccc}
A_1 & A_2& A_3&A_4
\end{array}\right) M^2_{A} \left(\begin{array}{cccc}
A_1 &
A_2 &
A_3 &
A_4
\end{array}\right)^T,
\eea where
 \bea
M^2_A = \left(\begin{array}{cccc}
 \fr{v^2(-2\lambda_{2S}u^2+2\lambda_{2S}v^2-\lambda_2\Lambda^2)}{2(u^2-v^2)} & -\fr{uv[2\lambda_{2S}(u^2-v^2)+\lambda_2\Lambda^2]}{2(u^2-v^2)} & \fr{uv^2[2\lambda_{2S}(u^2-v^2)+\lambda_2\Lambda^2]}{2(u^2-v^2)w} & 0\\
-\fr{uv[2\lambda_{2S}(u^2-v^2)+\lambda_2\Lambda^2]}{2(u^2-v^2)}& -\fr{u^2[2\lambda_{2S}(u^2-v^2)+\lambda_2\Lambda^2]}{2(u^2-v^2)} & \fr{u^2v[2\lambda_{2S}(u^2-v^2)+\lambda_2\Lambda^2]}{2(u^2-v^2)w}& 0\\
\fr{uv^2[2\lambda_{2S}(u^2-v^2)+\lambda_2\Lambda^2]}{2(u^2-v^2)w}&\fr{u^2v[2\lambda_{2S}(u^2-v^2)+\lambda_2\Lambda^2]}{2(u^2-v^2)w} & -\fr{u^2v^2[2\lambda_{2S}(u^2-v^2)+\lambda_2\Lambda^2]}{2(u^2-v^2)w^2} & 0\\
0 &0& 0 & 0
\end{array}\right).\label{gvh}
\eea
The above mass matrix provides only a combination of the pseudoscalars as a physical pseudoscalar, called $\mathcal{A}$, with
mass, $m^2_{\mathcal{A}}$, obtained by 
 \bea \mathcal{A} =\fr{vw A_1 + uw A_2 - u v A_3}{\sqrt{(u^2+v^2)w^2+u^2v^2}},\hs
 m^2_{\mathcal{A}} = - \fr{[v^2w^2+u^2(v^2+w^2)][2\lambda_{2S}(u^2-v^2)+\lambda_2\Lambda^2]}{2(u^2-v^2)w^2},\eea which is in $w,\La$ scales.
The remainders are three massless pseudoscalars,
\bea
G_Z = \fr{- uA_1 + v A_2}{\sqrt{u^2+v^2}},\hs G_{\mathcal{Z}_1}=A_4,\hs G_{\mathcal {Z}_1^\prime} =  \fr{uv^2A_1+u^2vA_2 + w(u^2+v^2) A_3}{\sqrt{(u^2+v^2)(u^2v^2+w^2u^2+w^2v^2)}},
\eea
which are the Goldstone bosons of the neutral gauge bosons, $Z,\ \mathcal {Z}_1$ and $\mathcal {Z}_1^\prime$, respectively.

For the charged scalar sector, $\Xi_{22}^{\pm \pm}$, $\Xi_{23}^{\pm (q-1)}$, and $\Xi_{33}^{\pm 2q}$ do not mix, and are physical fields by themselves
with masses,
\bea
m^2_{\Xi_{22}^{\pm \pm} } &=& \fr {\lambda_2(v^2-u^2)-2\lambda_{2\Xi}\Lambda^2}{2}, \\ 
m^2_{\Xi_{23}^{\pm (q-1)} } &=& \fr {\lambda_2(v^2-2u^2)+\lambda_3 w^2-4\lambda_{2\Xi}\Lambda^2}{4},\\
m^2_{\Xi_{33}^{\pm 2q}} &=& \fr {\lambda_3 w^2-\lambda_2 u^2-2\lambda_{2\Xi}\Lambda^2}{2},
\eea which are all in $w,\La$ scales. 
The remaining charged scalars mix in terms of 
\bea
 V^{\mathrm{charged}}_{\mathrm{mass}}&\supset& \left(\begin{array}{ccc}
S_{12}^+ & S_{21}^+& \Xi_{12}^+
\end{array}\right) M^2_{C_1} \left(\begin{array}{c}
S_{12}^- \\
S_{21}^-\\
\Xi_{12}^-
\end{array}\right)+ \left(\begin{array}{ccc}
S_{13}^{q} &\phi_1^{q}& \Xi_{13}^{q}
\end{array}\right) M^2_{C_q} \left(\begin{array}{c}
S_{13}^{-q} \\
\phi_1^{-q}\\
\Xi_{13}^{-q}
\end{array}\right)\crn
&&+ \left(\begin{array}{cc}
S_{23}^{(q+1)} & \phi_{2}^{(q+1)}
\end{array}\right) M^2_{C_{(q+1)}} \left(\begin{array}{c}
S_{23}^{-(q+1)} \\
\phi_{2}^{-(q+1)}
\end{array}\right),
\eea where $M^2_{C_1}$, $M^2_{C_q}$, and $M^2_{C_{(q+1)}}$ are mass-squared matrices for the singly, $q$, and $(q+1)$ charged scalars, respectively (as shown below). 

First, we derive 
\bea
M^2_{C_1} = \left(\begin{array}{ccc}
 \fr{\lambda_{2}u^2\Lambda^2}{2(v^2-u^2)} & \fr{\lambda_{2}uv\Lambda^2}{2(v^2-u^2)} & \fr{\lambda_{2}u\Lambda}{2\sqrt{2}} \\
\fr{\lambda_{2}uv\Lambda^2}{2(v^2-u^2)}& \fr{\lambda_{2}v^2\Lambda^2}{2(v^2-u^2)} &  \fr{\lambda_{2}u\Lambda}{2\sqrt{2}}\\
 \fr{\lambda_{2}u\Lambda}{2\sqrt{2}}& \fr{\lambda_{2}u\Lambda}{2\sqrt{2}} & \fr{\lambda_2(v^2-u^2)}{4}
\end{array}\right).\label{md1}
\eea
This leads to a physical, singly-charged field, with mass in the $w,\Lambda$ scales, 
 \bea H_5^\pm = \fr{\sqrt{2}u\Lambda S_{12}^\pm+\sqrt{2}v\Lambda S_{21}^\pm+(v^2-u^2)\Xi_{12}^\pm}{\sqrt{2(u^2+v^2)\Lambda^2+(v^2-u^2)^2}},\hs
 m^2_{H_5^\pm } = \fr {\lambda_2} {4} \left[v^2-u^2+\fr{2(u^2+v^2)\Lambda^2}{v^2-u^2}\right].\eea
Two remaining states are massless as combined of 
\bea G_{W_1}^\pm = \fr{- vS_{12}^\pm + u S_{21}^\pm}{\sqrt{u^2+v^2}}, \hs G_{W_2}^\pm= \fr{u(u^2-v^2)S_{12}^\pm+ v(u^2-v^2)S_{21}^\pm+\sqrt{2}(u^2+v^2)\Lambda\Xi_{12}^\pm}{\sqrt{(u^2-v^2)^2(u^2+v^2)+2(u^2+v^2)^2\Lambda^2}},
\eea
which are the Goldstone bosons of $W^\pm_1, W^\pm_2$ gauge bosons, respectively.

The mass matrix for $q$ charged scalars is 
\bea
M^2_{C_q} =\left(\begin{array}{ccc}
\frac{\sqrt{2}f v w}{u}+\frac{1}{2}(\la_1w^2-\la_2 \La^2) & \sqrt{2} f v+ \fr{\la_1 uw}{2}& \fr{\la_2 u \La}{2 \sqrt{2}} \\
\sqrt{2} f v+ \fr{\la_1 uw}{2} & \frac{1}{2} (\la_1 u^2+ \la_3 \La^2 +\fr{2\sqrt{2} f u v}{w}) & \fr{\la_3 w \La}{2 \sqrt{2}} \\
 \fr{\la_2 u \La}{2 \sqrt{2}}&\fr{\la_3 w \La}{2 \sqrt{2}} &\fr{1}{4}(\la_3 w^2-\la_2 u^2)
\end{array}\right).
\eea
We obtain a massless state,
$G_{X}^{\pm q}= \fr{uS_{13}^{\pm q}-w\phi_1^{\pm q} + \sqrt{2}\Lambda\Xi_{13}^{\pm q}}{\sqrt{u^2+w^2+2\Lambda^2}}$, as the Goldstone boson of $X^{\pm q}$ gauge boson. To find the remaining states, we define
\bea
H_6^{\prime \pm q} = \frac{ w S_{13}^{\pm q} +u \phi_1^{\pm q}}{\sqrt{u^2+w^2}}, \hs
H_7^{\prime \pm q}= \frac{-\sqrt{2} u \La S_{13}^{ \pm q} +\sqrt{2} w \La \phi_1^{\pm q} +(u^2+w^2) \Xi_{13}^{\pm q}}{\sqrt{(u^2+w^2)(u^2+w^2+2 \La^2)}},
\eea which are orthogonal to $G_{X}^{\pm q}$. The corresponding physical fields as the combinations of $H_{6}^{\prime \pm q}$, $H_7^{\prime \pm q}$ with masses are given by   
\bea
H_6^{\pm q} &=& c_{\varphi_q} H_6^{\prime \pm q} - s_{\varphi_q} H_7^{\prime \pm q}, \hs m^2_{H_6^{\pm q} } \simeq \fr {\lambda_1(u^2-v^2)w^2-\lambda_2u^2\Lambda^2}{2(u^2-v^2)},\crn
H_7^{\pm q} &=& s_{\varphi_q} H_6^{\prime \pm q}+c_{\varphi_q} H_7^{\prime \pm q},\hs  m^2_{H_7^{\pm q} } \simeq \fr {\lambda_3(w^2+2\Lambda^2)}{4},
\eea
where the $H_{6}^{\prime \pm q}$-$H_7^{\prime \pm q}$ mixing angle, called $\varphi_q$, is defined by
\bea t_{2\varphi_q} \simeq \fr{2(\la_2+\la_3)u\La\sqrt{2(\La^2+w^2)}}{-2\la_1w^3+\la_3w(w^2+2\La^2)+\fr{2\la_2wu^2\La^2}{u^2-v^2}},
\eea which is small due to $u,v\ll w,\La$, implying that such states slightly mix.   

Lastly, there remain two $(q+1)$ charged scalars. One of them is massless to be identified as the Goldstone boson of $Y^{\pm (q+1)}$ gauge boson, 
\bea
G_{Y}^{\pm (q+1)}= \fr{-vS_{23}^{\pm (q+1)} + w\phi_{2}^{\pm (q+1)}}{\sqrt{v^2+w^2}}.
\eea
The field that is orthogonal to it is heavy with mass in the $w, \La$ scales, given by  
 \bea H_8^{\pm (q+1)} &=& \fr{wS_{23}^{\pm (q+1)} + v\phi_{2}^{\pm (q+1)}}{\sqrt{v^2+w^2}},\crn
 m^2_{H_8^{\pm (q+1)}}& =& -\fr {(v^2+w^2)[(u^2-v^2)(2\lambda_{2S} u^2-\lambda_1 w^2)+\lambda_2u^2\Lambda^2]}{2(u^2-v^2)w^2}.
\eea

In summary, the model contains twelve massive Higgs fields, $H^0_{1,2,3,4}$, $\mathcal{A}^0$, $H^{\pm}_{5}$, $H^{\pm q}_{6,7}$, $H^{\pm (q+1)}_{8}$, $\Xi_{22}^{\pm\pm}$, $\Xi_{23}^{\pm (q-1)}$, and $\Xi_{33}^{\pm 2q}$, in which $H_1$ is the standard model like Higgs boson with mass in the weak scale, while the others are new, heavy Higgs bosons with masses in $w,\La$ scales. Besides, there are eleven massless Goldstone bosons, which
are correspondingly eaten by the eleven massive gauge bosons (where the conjugated fields are also counted). At the leading order, the physical scalar states are related to those in the gauge basis as   
\bea
&&\left(   \begin{array}{c}     H_1 \\     H_2 \\   \end{array} \right) \simeq  \left(   \begin{array}{cc}     c_{\alpha_1} & s_{\alpha_1} \\     -s_{\alpha_1} & c_{\alpha_1} \\   \end{array}\right) \left(   \begin{array}{c}    S_1 \\    S_2 \\  \end{array}\right),\hs \left(   \begin{array}{c}    H_3 \\    H_4 \\  \end{array}\right) \simeq \left(  \begin{array}{cc}  c_\varphi & -s_\varphi \\  s_\varphi & c_\varphi\\  \end{array} \right)\left(  \begin{array}{c}    S_3 \\    S_4 \\  \end{array}\right),\crn
&&\left(  \begin{array}{c}   \mathcal{A} \\   G_Z \\ G_{\mathcal{Z}_1} \\ G_{\mathcal{Z}_1^\prime} \\ \end{array}\right) \simeq  \left(\begin{array}{cccc}s_{\alpha_1} & c_{\alpha_1} &-\fr{u}{w}s_{\alpha_1}&0  \\  -c_{\alpha_1} &  s_{\alpha_1} &0&0 \\ 0&0&0&1 \\ \fr{v}{2w}s_{2\alpha_1}&\fr{u}{2w}s_{2\alpha_1}&1&0  \\  \end{array}\right)\left(  \begin{array}{c}    A_1 \\    A_2 \\  A_3 \\  A_4 \\  \end{array}\right), \crn
&&\left(  \begin{array}{c}  H_5^\pm \\  G_{W_1}^\pm \\ G_{W_2}^\pm \\ \end{array}\right) \simeq  \left(\begin{array}{ccc}c_{\alpha_1} & s_{\alpha_1} &\fr{v^2-u^2}{\sqrt{2}\sqrt{u^2+v^2}\Lambda}\\  -s_{\alpha_1} &  c_{\alpha_1} &0 \\  \fr{u}{\sqrt{2}\Lambda}c_{2\alpha_1}&\fr{v}{\sqrt{2}\Lambda}c_{2\alpha_1}&1  \\  \end{array}\right)\left(  \begin{array}{c}     S_{12}^\pm \\   S_{21}^\pm \\  \Xi_{12}^\pm \\  \end{array}\right), \label{lhh} \\
&&\left(  \begin{array}{c}  G_{X}^{\pm q} \\  H_6^{\pm q} \\ H_7^{\pm q} \\ \end{array}\right) \simeq  \left(\begin{array}{ccc} \fr{u}{w}s_{\al_2} & -s_{\alpha_2} & c_{\alpha_2}  \\ c_{\varphi_q}& \fr{u}{w}c_{\varphi_q}-  c_{\alpha_2}s_{\varphi_q} & - s_{\alpha_2}s_{\varphi_q} \\  s_{\varphi_q}-\fr{u}{w} c_{\alpha_2}c_{\varphi_q} & c_{\alpha_2}c_{\varphi_q} & s_{\alpha_2}c_{\varphi_q}  \\  \end{array}\right)\left(  \begin{array}{c}     S_{13}^{\pm q} \\   \phi_{1}^{\pm q}  \\  \Xi_{13}^{\pm q} \\  \end{array}\right), \crn
&&\left(  \begin{array}{c}  G_{Y}^{\pm(q+1)} \\  H_{8}^{\pm(q+1)} \\ \end{array}\right) \simeq  \left(\begin{array}{cc} -\fr{v}{w} &1 \\ 1 & \fr{v}{w} \\  \end{array}\right)\left(  \begin{array}{c}     S_{23}^{\pm (q+1)} \\   \phi_{2}^{\pm (q+1)}  \\  \end{array}\right),\nn 
\eea
where the $\al_{1,2}$ angles have been introduced, defined by $t_{\alpha_1}=v/u$ and $t_{\alpha_2}=w/\sqrt{2}\Lambda$, respectively.

\section{Gauge sector\label{gaugesec}}

Let us investigate the mass spectrum of the gauge bosons in the considering model. When the scalars develop the VEVs, the gauge bosons get masses as derived from
\bea\mathcal{L}_{s} = \Tr [(D_\mu  S)^\dag (D^\mu S)+(D_\mu \Xi)^\dag (D^\mu \Xi)]+(D_\mu \phi)^\dag (D^\mu \phi),
\label{gma1}\eea 
where the covariant derivatives are defined by
\bea D_\mu S &=& \partial_\mu S+ig_L \frac{ \sigma_a}{2} A_{aL \mu} S -ig_R S \frac{ \lambda_i}{2} A_{iR \mu}+i g_X X_S B_\mu S,  \\
D_\mu \Xi &=&\partial_\mu \Xi+ ig_R \frac{ \lambda_i}{2} A_{iR \mu} \Xi + i g_R \Xi \frac{ \lambda^*_i }{2} A_{iR \mu}  +i g_X X_{\Xi} B_\mu \Xi, \\
D_\mu \phi &=& \partial_\mu \phi +ig_R \frac{ \lambda_i}{2} A_{iR \mu}  \phi+i g_X X_{\phi} B_\mu \phi.\label{dhhb}\eea
Here, $\sigma_a$ are the Pauli matrices, and $\lambda_i$ are the Gell-Mann matrices. $g_L$, $g_R$, and $g_X$ are the gauge coupling constants of $SU(2)_L, SU(3)_R$, and $U(1)_X$, respectively. $X_{S, \Xi,  \phi}$ stand for
the $U(1)_X$ charges of the corresponding scalar multiplets.  

Substituting the VEVs for $S, \Xi,  \phi$, we obtain the mass Lagrangian,
\bea
\mathcal{L}^{\mathrm{gauge}}_{\mathrm{mass}}
 &=&\frac{g^2_L}{8}  \left \{ \left[ A_{3L}^\mu - t_R A_{3R}^\mu -\frac{t_R}{\sqrt{3}}  A_{8R}^\mu -\frac{ (1+2q) t_X}{3} B^\mu \right]^2 +2 \left[ W_L^{\mu+} W_{L\mu}^- 
 +t_R^2 ( W_R^{\mu +} W_{ R\mu}^- \right.\right.\crn
 &&\left.\left.+ X_{R \mu}^q X_{R}^{-q\mu})\right] \right\} u^2 + \frac{g^2_L}{8}  \left \{ \left[ A_{3L}^\mu - t_R A_{3R}^\mu +\frac{t_R}{\sqrt{3}}  A_{8R}^\mu +\frac{ (1+2q) t_X}{3} B^\mu \right]^2 +2 \left[ W_L^{\mu+} \right.\right.\crn
 &&\left.\left.\times W_{L\mu}^- +t_R^2 \left( W_R^{\mu +} W_{ R\mu}^- + Y_{R \mu}^{q+1} Y_{R}^{ -(q+1)\mu}\right)\right] \right\} v^2 - \frac{g_L^2}{2}\left( W_{L \mu}^- W_{R}^{\mu +}+ W_{L \mu}^+ W_R^{\mu -} \right)t_R u v\crn
 && +
\frac{g_L^2}{2}\left\{ \left[ t_R A_{3R}^\mu +\frac{t_R}{\sqrt{3}} A_{8R}^\mu +\frac{2 t_X}{3} (q-1) B^\mu\right]^2+t_R^2\left(W_{R \mu}^+ W_{R} ^{\mu -} +X_{R \mu}^q  X_{R}^{ -q \mu}\right)\right \}\Lambda^2 \crn
&& + \frac{g_L^2}{18}\left\{ \left[\sqrt{3} t_R A_{8R}^\mu+t_X(1+2q)B^\mu\right]^2+\frac{9}{2}t_R^2 \left(X_{R \mu}^q  X_{R}^{ -q \mu}+Y_{R \mu}^{q+1} Y_{R}^{-(q+1)\mu }\right)\right \}w^2, \crn
&=&\frac{g_L^2 t_R^2}{4} (v^2+w^2) Y_{R \mu}^{-(q+1)}Y_R^{ (q+1)\mu } +  \frac{g_L^2 t_R^2}{4} (u^2+w^2+2\Lambda^2) X_{R\mu}^{-q} X_R^{q \mu}+ (W_{L}^{\mu +}\ W_{R}^{\mu +})\crn
&&\times  M_{W}^2 \left(W_{ L\mu }^-\ W_{R \mu}^-\right)^T+  \frac{1}{2}(A_{3L}^\mu\ A_{3R}^\mu\ A_{8R}^\mu\ B^\mu) M^2_0 \left(A_{3L \mu}\ A_{3R \mu}\ A_{8R \mu}\ B_\mu\right)^T,
\label{mass}
\eea
where we have denoted, $t_X=\frac{g_X}{g_{L}}$, $t_{R} =\frac{g_R}{g_L}$, and the non-Hermitian gauge bosons as
\bea
&& W_{L\mu}^{\pm} =\frac{1}{\sqrt{2}} \left(A_{1L \mu} \mp i A_{2L \mu} \right), \hs \hs \hs W_{R\mu}^{\pm} =\frac{1}{\sqrt{2}} \left(A_{1R \mu} \mp i A_{2R \mu} \right),  \label{bxddh1}\\ && Y_{R\mu}^{\pm (q+1)} =\frac{1}{\sqrt{2}} \left(A_{6R \mu} \pm i A_{7R \mu} \right),  \hs \hs X_{R\mu}^{ \pm q} =\frac{1}{\sqrt{2}} \left(A_{4R \mu} \pm i A_{5R \mu} \right).\label{gaugec1}
\eea
The mass Lagrangian in (\ref{mass}) has be rewritten in terms of the matrix forms, where $M_{W}$ and $M_0$ define the mass matrices of the left-right $W$ and neutral gauge bosons, respectively.   

We see that the gauge bosons, $X_{R\mu}^{\pm q}$ and $Y_{R\mu}^{\pm(q+1)}$, by themselves are physical with masses,
\bea
m^2_{X_R}= \frac{g_R^2}{4}  (u^2+w^2+2\Lambda^2), \hs m^2_{Y_R} =\frac{g_R^2}{4} (v^2+w^2).
\eea
The left-right $W$ bosons mix via a mass matrix as given by 
\bea
M_{W}^  2= \frac{g_L^2}{4} \left(%
\begin{array}{ccc}
  u^2 +v^2  & -2t_R u v \\
  - 2t_R u v & t_R^2 (u^2+v^2+2 \Lambda^2) \\
\end{array}%
\right).
\eea
Diagonalizing this matrix, we obtain two physical states,
\bea
W_{1 \mu}^\pm = c_\xi W_{L \mu}^\pm -s_\xi W_{R \mu}^\pm, \hs  W_{2 \mu}^\pm = s_\xi W_{L \mu}^\pm + c_\xi W_{R \mu}^\pm,\label{gaugec2}
\eea
where the $W_L$-$W_R$ mixing angle ($\xi$) is obtained by $t_{2\xi}=\tan 2\xi = \frac{-4 t_R u v}{2t_R^2 \Lambda^2 +(t_R^2-1)(u^2+v^2) }$. And, the corresponding masses are 
\bea
m_{W_1}^2 & \simeq& \frac{g_L^2}{4} \left[ u^2+v^2 - \frac{ 4t_R^2 u^2 v^2}{2t_R^2 \Lambda^2 +(t_R^2-1)(u^2+v^2) }\right],  \crn m_{W_2}^2 & \simeq& \frac{g_R^2}{4} \left[ u^2+v^2+2 \Lambda^2 + \frac{ 4 u^2 v^2}{2t_R^2 \Lambda^2 +(t_R^2-1)(u^2+v^2) }\right].
\eea
Because of the condition, $u,v\ll w,\La$, the $W_1$ boson has a small mass in the weak scales ($u,v$) which is identical to the standard model $W$ boson, whereas the $W_2$ boson is a new, heavy charged gauge boson with the mass proportional to  
$\La$ scale. The mixing between these two fields is small since $\xi\rightarrow 0$ due to the above condition.  

The diagonalization of the neutral gauge boson sector is more complicated, because all the four gauge fields generally mix. Indeed, the mass matrix is given by
\bea
 M^2_0=\frac{g_L^2}{4} \left(%
\begin{array}{cccc}
  u^2 +v^2  & -t_R (u^2+v^2)&-\frac{t_R}{\sqrt{3}}(u^2-v^2)&  \frac{\beta t_X}{\sqrt{3}}(u^2-v^2)\\
 -t_R (u^2+v^2) & t_R^2 (u^2+v^2+4 \Lambda^2)& \frac{t_R^2}{\sqrt{3}}(u^2-v^2+4\Lambda^2) 
 & m^2_{42} \\
 -\frac{t_R}{\sqrt{3}}(u^2-v^2)&\frac{t_R^2}{\sqrt{3}}(u^2-v^2+4\Lambda^2)&\frac{t_R^2}{3}[u^2+v^2+4(w^2+\Lambda^2)]&
 m_{43}^2 \\
 \frac{\beta t_X}{\sqrt{3}}(u^2-v^2) &m^2_{42}& m^2_{43}& m^2_{44}
\end{array}%
\right),
\eea
where 
\bea
 m^2_{42} &=& -\frac{t_R t_X}{3} [\sqrt{3}\beta (u^2-v^2+4 \Lambda^2)+12 \Lambda^2], \crn  m^2_{43} &=& \frac{-t_R t_X}{3}[\beta(u^2+v^2+4w^2+4 \Lambda^2)+4 \sqrt{3} \Lambda^2],  \crn  m^2_{44} &=& \frac{t_X^2}{3} [(u^2+v^2+4 w^2) \beta^2+4(\sqrt{3}+\beta)^2\Lambda^2]. \nn
\eea

First of all, from the mass matrix, we can always obtain a zero eigenvalue (i.e. photon mass) with the corresponding eigenstate (i.e. photon field)
as 
\bea
A_{\mu}=\frac{t_R t_X}{\sqrt{t_R^2+t_X^2(1+\beta^2+t_R^2)}}\left(A_{3L \mu}+ \frac{1}{t_R}A_{3R \mu}+\frac{\beta}{t_R}A_{8R \mu}+\frac{1}{t_X}B_\mu \right),
\eea
which is independent of the VEVs as a consequence of the electric charge 
conservation \cite{chargequan}. Next, we can determine electromagnetic interactions following the standard procedure in \cite{chargequan}, and thus the Weinberg's angle ($\theta_W$) is identified as
\bea
s_W= \frac{t_R t_X}{\sqrt{t_R^2+t_X^2(1+\beta^2+t_R^2)}},
\eea
where note that $s_W=\sin\theta_W$, $c_W=\cos \theta_W$, and so forth. With this at hand, the photon field is rewritten in terms of 
\bea
A_\mu = s_W A_{3L \mu}+c_W \left(\frac{t_W}{t_R}A_{3R \mu}+\beta  \frac{t_W}{t_R} A_{8R \mu}+ \frac{t_W}{t_X} B_\mu \right),
\eea where the parentheses present the field as coupled to the weak hypercharge $Y=T_{3R}+\beta T_{8R}+X$.     

The standard model $Z$ boson is orthogonal to the photon field as usual,  
\bea
Z_\mu = c_W A_{3L \mu}-s_W \left(\frac{t_W}{t_R}A_{3R \mu}+\beta  \frac{t_W}{t_R} A_{8R \mu }+ \frac{t_W}{t_X} B_\mu\right).
\eea
The model under consideration contains two new neutral gauge bosons, called $Z_{R }$ and $Z_{R }^\prime$, which are given orthogonally to the hypercharge field in the parentheses (i.e., orthogonal to both the photon and $Z$ fields). Thus, they are obtained by  
\bea
Z'_{R \mu} &=& \frac{1}{\sqrt{t_R^2+\beta^2 t_X^2}} \left( t_R A_{8R \mu}- 
 \beta t_X  B_{\mu} \right),\\
Z_{R \mu} &=& \frac{1}{\sqrt{(t_R^2+\beta^2 t_X^2)[t_R^2+(1+\beta^2)t_X^2]}} \left[-(t_R^2+\beta^2 t_X^2) A_{3R \mu}+\beta t_X^2 A_{8R \mu}+t_X t_R B_\mu \right], 
\eea
where note that $t_X= \frac{s_W t_R}{\sqrt{t_R^2-(1+\beta^2+t_R^2)s_W^2}}$, and these new states must be heavy. 

Next, let us change to the new basis consisting of $A_\mu$, $Z_\mu$, $Z^{\prime}_{R\mu}$, and $Z_{R\mu}$ by the transformation,   
$(A_{3L \mu}\ A_{3R \mu}\ A_{8R \mu}\ B_\mu)^T=U(A_\mu\ Z_\mu\ Z'_{R \mu}\ Z_{R \mu})^T$, 
where 
\bea
U= \left(%
\begin{array}{cccc}
s_W&c_W&0&0\\
\frac{s_W}{t_R}&-\frac{s_W t_W}{t_R}&0&
-\frac{t_R^2+\beta^2 t_X^2}{\sqrt{(t_R^2+\beta^2 t_X^2)[t_R^2+(1+\beta^2)t_X^2]}}\\
\frac{\beta s_W}{t_R}&-\frac{\beta s_W t_W}{t_R}&\frac{t_R}{\sqrt{t_R^2+\beta^2 t_X^2}}&\frac{\beta t_X^2}{\sqrt{(t_R^2+\beta^2 t_X^2)[t_R^2+(1+\beta^2)t_X^2]}}\\
\frac{s_W}{t_X}&-\frac{s_W t_W}{t_X}&-\frac{\beta t_X}{\sqrt{t_R^2+\beta^2 t_X^2}}&\frac{t_X t_R}{\sqrt{(t_R^2+\beta^2 t_X^2)[t_R^2+(1+\beta^2)t_X^2]}}
\end{array} \right).
\eea Correspondingly, the mass matrix $M_0^2$ is changed to 
\bea
 M^{\prime 2}_0 =U^T M^2_0 U =\left(%
\begin{array}{ccc} 
0& 0 \\
0& M^{\prime 2}
\end{array} \right). 
\eea
 We see that the photon field, $A_\mu$, is decoupled as a physical massless field, while the other states $(Z_\mu, Z'_{R\mu}, Z_{R\mu})$ mix by themselves via a $3 \times 3$ mass matrix found to be 
\bea
M^{\prime 2}=\frac{g_L^2}{4} \left(%
\begin{array}{ccc}
\frac{u^2+v^2}{c_W^2} & -\frac{(u^2-v^2)\kappa c_W}{\sqrt{3}[t_R^2+t_X^2(1+\beta^2)]}
 & \frac{t_R(u^2+v^2)\kappa c_W}{[t_R^2+t_X^2(1+\beta^2)]^{3/2}}  \\
 -\frac{(u^2-v^2)\kappa c_W}{\sqrt{3}[t_R^2+t_X^2(1+\beta^2)]} &
  \frac{(t_R^2+\beta^2 t_X^2)^2(u^2+v^2+4w^2)+4 \kappa^\prime \Lambda^2}{3(t_R^2+\beta^2 t_X^2)}
   & \frac{\kappa^{\prime \prime}(v^2-u^2)}{\sqrt{3}}-\frac{4 t_R^2 \sqrt{\kappa'}\Lambda^2}{\sqrt{3}\kappa^{\prime \prime} }  \\
 \frac{t_R(u^2+v^2)\kappa c_W}{[t_R^2+t_X^2(1+\beta^2)]^{3/2}}
 & \frac{\kappa^{\prime \prime}(v^2-u^2)}{\sqrt{3}}-\frac{4 t_R^2 \sqrt{\kappa'}\Lambda^2}{\sqrt{3}\kappa^{\prime \prime} }
 & t_R^2\left(\frac{u^2+v^2}{\kappa^{\prime \prime \prime}}+4\kappa^{\prime \prime \prime}\Lambda^2\right)  \\
\end{array}%
\right),  \nonumber    \label{gauge1} \eea 
where we have conveniently denoted, 
\bea && \kappa  = [t_R^2(1+t_X^2)+(1+\beta^2) t_X^2] \sqrt{t_R^2+\beta^2
t_X^2}, \hs  \kappa^\prime = [t_R^2+(\sqrt{3}+\beta)\beta t_X^2]^2,\nonumber \\
&& \kappa^{\prime \prime} = t_R(t_R^2+\beta^2
t_X^2)/\sqrt{t_R^2+(1+\beta^2)t_X^2}, \hs \kappa^{\prime \prime \prime}=[t_R^2+(1+\beta^2)t_X^2]/(t_R^2+\beta^2 t_X^2). \eea

Because of the condition, $u,v \ll w, \Lambda$, the first row and first column of $M^{\prime 2}$ consist of the elements that are 
much smaller than those of the remaining entries. Hence, we can diagonalize $M^{\prime 2}$ by using the familiar seesaw formula.  We introduce a  basis ($\mathcal{Z}_\mu, \mathcal{Z}^\prime_{R \mu},\mathcal{Z}_{R \mu}$) in such a way as to
 separate the light $\mathcal{Z}_\mu$ boson from the two heavy  $\mathcal{Z}^\prime_{R \mu} ,\mathcal{ Z}_{R \mu}$ bosons. This
 basis is related to the previous basis ($Z_\mu, Z^\prime_{R \mu}, Z_{R \mu}$) by an unitary transformation  as $( Z_\mu\ Z^\prime_{R \mu}\ Z_{R \mu})^T= \mathcal{U} ( \mathcal{Z}_\mu\  \mathcal{Z}^\prime_{R \mu}\ \mathcal{ Z}_{R \mu})^T$. Correspondingly, the mass matrix, $M^{\prime 2}$, is changed to  
\bea
\mathcal{M^{\prime }}^{2} = \mathcal{U}^T M^{\prime 2} \mathcal{U} = \left(%
\begin{array}{cc} m^2_{\mathcal{Z}} & 0 \\
       0 & \mathcal{M}_{2 \times 2}^2\\
\end{array}%
\right).
\eea
Using the seesaw approximation, we obtain
 \bea
 && \mathcal{U}\simeq  \left(%
\begin{array}{ccc}
1 & \epsilon_1 & \epsilon_2 \\
-\epsilon_1 & 1 & 0 \\
-\epsilon_2 &0 &1
 \end{array}%
\right),\crn
&& m^2_{\mathcal{Z}}\simeq \frac{g_L^2}{4}\left\{ \frac{u^2+v^2}{c_W^2}+  \frac{\epsilon_1(u^2-v^2)\kappa c_W}{\sqrt{3}[t_R^2+t_X^2(1+\beta^2)]}
- \frac{\epsilon_2 t_R^2(u^2+v^2)\kappa c_W}{[t_R^2+t_X^2(1+\beta^2)]^{\frac{3}{2}}}\right\},\crn 
&& \mathcal{M}_{2 \times 2}^2 \simeq \frac{g_L^2}{4}\left(%
\begin{array}{cc}
\frac{(t_R^2+\beta^2 t_X^2)^2(u^2+v^2+4w^2)+4 \kappa^\prime \Lambda^2}{3(t_R^2+\beta^2 t_X^2)}
   & \frac{\kappa^{\prime \prime}(v^2-u^2)}{\sqrt{3}}-\frac{4t_R^2\sqrt{\kappa'}\Lambda^2}{\sqrt{3}\kappa^{\prime \prime}} \\
 \frac{\kappa^{\prime \prime}(v^2-u^2)}{\sqrt{3}}-\frac{4t_R^2\sqrt{\kappa'}\Lambda^2}{\sqrt{3}\kappa^{\prime \prime}}
 & t_R^2\left(\frac{u^2+v^2}{\kappa^{\prime \prime \prime}}+4\kappa^{\prime \prime \prime}\Lambda^2\right)  \\
 \end{array}%
\right), \nn
\label{gaugeq} \eea
where $\epsilon_{1,2}$ are defined as
\bea
\epsilon_1  & = &  \frac{\sqrt{3}\kappa c_W}{4\kappa^{\prime \prime \prime}}\left\{ \frac{-(u^2+v^2)}{[ t_R^2+\beta t_X^2(\sqrt{3}+ \beta)][t_R^2+(1+\beta^2)t_X^2] \Lambda^2}\right.\crn
&&\left. - \frac{(u^2-v^2)}{(t_R^2+\beta^2 t_X^2)^2w^2+[t_R^2+(\sqrt{3}+\beta)\beta t_X^2]^2\Lambda^2}\right\}, \\
 \epsilon_2  & = &  \frac{\kappa c_W}{4\kappa^{\prime \prime}} \left\{  \frac{u^2-v^2}{t_R[t_R^2+\beta t_X^2(\sqrt{3}+\beta)][t_R^2+t_X^2(1+\beta^2)] \Lambda^2}\right.\crn
 &&\left. +\frac{u^2+v^2}{t_R[t_R^2+(1+\beta^2)t_X^2]^2 \Lambda^2} \right\}.
\eea
Note that $ \mathcal{M}_{2 \times 2}^2$ describes two heavy states, $\mathcal{Z}_R\simeq Z_R$ and $\mathcal{Z}'_R\simeq Z'_R$, as given at the leading order. The mixing between $Z$ and these heavy states is very suppressed, $\epsilon_1 ,\epsilon_2\ll 1$, due to $u,v\ll w,\La$. The $\mathcal{Z}_\mu$ boson is identical to 
the standard model $Z$ boson with mass, $ 
m^2_{\mathcal{Z}} \simeq  \frac{g_L^2}{4c_W^2}(u^2+v^2)$.

Finally, the states $\mathcal{Z}^\prime_R$ and $\mathcal{Z}_{R}$ still mix. Diagonalizing their mass matrix,
we obtain the corresponding physical states 
\bea
\mathcal{Z}_{1} = c_\epsilon \mathcal{Z}^\prime_R - s_\epsilon \mathcal{Z}_R, \hs \mathcal{Z}_{1}^\prime = s_\epsilon \mathcal{Z}^\prime_R + c_\epsilon \mathcal{Z}_R, \label{nnnn}
\eea
with masses, 
\bea
m^2_{\mathcal{Z}_{1}} & \simeq&\frac{g_L^2}{6}\left \{t_R^2(w^2+4 \Lambda^2)+t_X^2[\beta^2 w^2+(\sqrt{3}+\beta)^2\Lambda^2]   \right.  \nonumber \\  
 &&-\left. \sqrt{[t_R^2(w^2+4 \Lambda^2)+t_X^2(\beta^2 w^2+(\sqrt{3}+\beta)^2\Lambda^2)]^2-12t_R^2[t_R^2+(1+\beta^2)t_R^2]w^2 \Lambda^2} \right \}, \\
m^2_{\mathcal{Z}_{1}^\prime} &\simeq&\frac{g_L^2}{6}\left \{t_R^2(w^2+4 \Lambda^2)+t_X^2[\beta^2 w^2+(\sqrt{3}+\beta)^2\Lambda^2]   \right.  \nonumber \\  
 &&+\left. \sqrt{[t_R^2(w^2+4 \Lambda^2)+t_X^2(\beta^2 w^2+(\sqrt{3}+\beta)^2\Lambda^2)]^2-12t_R^2[t_R^2+(1+\beta^2)t_R^2]w^2 \Lambda^2} \right \}, 
\eea which are all in $w,\La$ scales. 
Above, the $\mathcal{Z}'_R$-$\mathcal{Z}_R$ mixing angle, $\epsilon$, is obtained by 
 \bea t_{2\epsilon} \simeq \frac{2 \sqrt{3}t_R^2(t_
R^2+\beta^2 t_X^2)[t_R^2+\beta(\beta+\sqrt{3})t_X^2]\Lambda^2}{\kappa^{\prime \prime}\left\{(t_R^2+\beta^2 t_X^2)^2 w^2-[2t_R^4+(\sqrt{3}-\beta)^2 t_R^2 t_X^2-(\sqrt{3}+\beta)^2\beta^2t_X^4]\Lambda^2\right\}}, \eea which is generally finite due to $w\sim \La$. 

To summarize, the physical neutral gauge bosons are related to the gauge states as $(A_{3L} \,\, A_{3R}\,\, A_{8R}\,\, B)^T=V(A\,\, \mathcal{Z}\,\, \mathcal{Z}_{1}\,\, \mathcal{Z}^\prime_{1})^T$, with
\bea V=U\mathcal{U}U_\epsilon \simeq U U_\epsilon= \left(
\begin{array}{cccc}
s_W & c_W & 0 & 0\\
\frac{s_W}{t_R}&-\frac{s^2_W}{t_Rc_W}&\frac{\sqrt{t_R^2+ t_X^2\beta^2}s_\epsilon s_W}{t_Rt_Xc_W}&-\frac{\sqrt{t_R^2+ t_X^2\beta^2}c_\epsilon s_W}{t_Rt_Xc_W}\\
\frac{\beta s_W}{t_R}&-\frac{\beta s^2_W}{t_Rc_W}&\frac{t_R^2c_\epsilon c_W -\beta t_Xs_\epsilon s_W}{t_Rc_W\sqrt{t_R^2+ t_X^2\beta^2}}&\frac{t_R^2s_\epsilon c_W + \beta t_Xc_\epsilon s_W}{t_Rc_W\sqrt{t_R^2+ t_X^2\beta^2}}\\
\frac{s_W}{t_X}&-\frac{s^2_W}{t_Xc_W}& \frac{-\beta t_Xc_\epsilon c_W -s_\epsilon s_W}{c_W\sqrt{t_R^2+ t_X^2\beta^2}}&\frac{-\beta t_Xs_\epsilon c_W + c_\epsilon s_W}{c_W\sqrt{t_R^2+ t_X^2\beta^2}}
\end{array} \right),\label{gaugen}\eea 
where $\mathcal{U}\simeq 1$ due to $\epsilon_{1,2}\ll 1$, and 
\bea U_\epsilon= \left(
\begin{array}{cccc}
1 & 0 & 0 & 0\\
0 &1& 0 & 0\\
0 & 0 &c_\epsilon &s_\epsilon \\
0&0& -s_\epsilon & c_\epsilon 
\end{array} \right).\eea 
For the following calculations, we will use $V$ as approximated, thus $\mathcal{Z}=Z$, and $\mathcal{Z}_1,\ \mathcal{Z}'_1$ are directly related to $Z'_R$, $Z_R$ by an expression like (\ref{nnnn}) since $\mathcal{Z}_R=Z_R$ and $\mathcal{Z}'_R=Z'_R$.

 \section{INTERACTIONS\label{interactionsec}}
 \subsection{Fermion--gauge boson interactions}
  
The gauge interactions of fermions arise from the Lagrangian,
  \bea
  \mathcal{L}_{f} &=&\bar{\Psi } i \gamma^\mu D_\mu \Psi  
  = \bar{\Psi }i \gamma^\mu \partial_\mu \Psi -g_L \bar{\Psi }_L \gamma^\mu (P^{CC}_{L\mu}+P^{NC}_{L\mu})\Psi_L - g_R \bar{\Psi}_R \gamma^\mu (P^{CC}_{R\mu}+P^{NC}_{R\mu}) \Psi_R,
\label{fh2}  \eea
where the covariant derivative is $D_\mu = \pa_\mu + i g_L T_{aL}A_{aL\mu}+ig_R T_{iR}A_{iR\mu}+ig_X X B_\mu$ and the gauge vectors relevant to the charged and neutral currents are obtained as   
\bea
P^{CC}_L&=&T_{1L}A_{1L}+T_{2L}A_{2L}, \hs P^{NC}_L=T_{3L}A_{3L}+t_X X_{\Psi_L}B,\crn
P^{CC}_R&=&\sum_{i=1,2,4,5,6,7} T_{iR}A_{iR},\hs
P^{NC}_R=T_{3R}A_{3R}+T_{8R}A_{8R}+\fr{t_X}{t_R} X_{\Psi_R}B.\nn
\eea
Above, $\Psi_L$ and $\Psi_R$ run on all the left-handed and right-handed fermion multiplets of the model, respectively. Note also that the interactions of fermions with gluons have the common form, which are easily determined and thus have been omitted.
  
Using (\ref{bxddh1}) and (\ref{gaugec1}) as well as (\ref{gaugec2}) for (\ref{fh2}), we derive the interactions of the physical charged gauge bosons with fermions as
\bea
\mathcal{L}_{CC} &=& -g_L\bar{\Psi }_L \gamma^\mu P^{CC}_{L\mu}\Psi_L - g_R \bar{\Psi}_R \gamma^\mu P^{CC}_{R\mu}\Psi_R\crn
&=&J_{1W}^{-\mu}W_{1\mu}^+ +J_{2W}^{-\mu}W_{2\mu}^+ +J_{X}^{-q\mu}X_{R\mu}^q +J_{Y}^{-(q+1)\mu}Y_{R\mu}^{q+1}+H.c.,
\eea
where the charged currents $J_{1W}^{-\mu}$, $J_{2W}^{-\mu}$, $J_{X}^{-q\mu}$, and $J_{Y}^{-(q+1)\mu}$ are respectively defined as 
\bea
&& J_{1W}^{-\mu}=-\fr{g_Lc_\xi }{\sqrt{2}} (\bar{\nu}_{aL}\gamma^\mu e_{aL}+\bar{u}_{aL}\gamma^\mu d_{aL})+\fr{g_R s_\xi }{\sqrt{2}}(\bar{\nu}_{aR}\gamma^\mu e_{aR}+\bar{u}_{aR}\gamma^\mu d_{aR}),\crn
&& J_{2W}^{-\mu}= -\fr{g_Ls_\xi}{\sqrt{2}} (\bar{\nu}_{aL}\gamma^\mu e_{aL}+\bar{u}_{aL}\gamma^\mu d_{aL})-\fr{g_Rc_\xi }{\sqrt{2}}(\bar{\nu}_{aR}\gamma^\mu e_{aR}+\bar{u}_{aR}\gamma^\mu d_{aR}),\crn
&& J_{X}^{-q\mu}=-\fr{g_R}{\sqrt{2}}(\bar{E}_{a R}\gamma^\mu \nu_{a R}-\bar{d}_{\alpha R}\gamma^\mu J_{\alpha R} +\bar{J}_{3 R}\gamma^\mu u_{3 R}),\crn
&& J_{Y}^{-(q+1)\mu}=-\fr{g_R}{\sqrt{2}}(\bar{E}_{a R}\gamma^\mu e_{a R} +\bar{u}_{\alpha R}\gamma^\mu J_{\alpha R} + \bar{J}_{3 R}\gamma^\mu d_{3 R}).
\eea

Using the physical neutral gauge bosons defined by (\ref{gaugen}), $P^{NC}_{L\mu}$ and $P^{NC}_{R\mu}$ become 

\bea
P^{NC}_{L\mu}&= &\, s_WQ_{\Psi_L}A_\mu+\fr {1}{c_W}(T_{3L}-s_W^2 Q_{\Psi_L}) Z_\mu+\fr{t_X (T_{3L}-Q_{\Psi_L})}{c_W\sqrt{t_R^2+ t_X^2\beta^2}}\crn
&&\times \left[ (\beta t_Xc_\epsilon c_W + s_\epsilon s_W)\mathcal{Z}_{1\mu}+(\beta t_Xs_\epsilon c_W - c_\epsilon s_W) \mathcal{Z}_{1\mu}^\prime\right],\crn
t_R P^{NC}_{R\mu}&= &s_WQ_{\Psi_R} A_\mu-\fr{1}{c_W}s^2_WQ_{\Psi_R}Z_\mu\crn
&&+\left\{\frac{c_W(t_X\beta c_\epsilon t_W+t_R^2s_\epsilon)T_{3R}-t_W(\beta t_Xc_\epsilon c_W +s_\epsilon s_W)Q_{\Psi_R}}{t_X^{-1}s_W\sqrt{t^2_R+t_X^2\beta^2}}+c_\epsilon \sqrt{t_R^2+t_X^2\beta^2} T_{8R}\right\} \mathcal{Z}_{1\mu}\crn
&&+\left\{\frac{c_W(t_X\beta s_\epsilon t_W-t_R^2c_\epsilon)T_{3R}-t_W(\beta t_Xs_\epsilon c_W -c_\epsilon s_W)Q_{\Psi_R}}{t_X^{-1}s_W\sqrt{t^2_R+t_X^2\beta^2}}+s_\epsilon\sqrt{t_R^2+t_X^2\beta^2} T_{8R}\right\} \mathcal{Z}^\prime_{1\mu},\nn\label{sccs}
\eea
where note that $Q_{\Psi_L}=T_{3L}+X_{\Psi_L}$ and $Q_{\Psi_R}=T_{3R}+\beta T_{8R}+X_{\Psi_R}$.  

Hence, we have the neutral current interactions from (\ref{fh2}) such that
\bea
\mathcal{L}_{NC}&=&-g_L \bar{\Psi }_L \gamma^\mu P^{NC}_{L\mu}\Psi_L - g_R\bar{\Psi}_R \gamma^\mu P^{NC}_{R\mu}\Psi_R\crn
&=&-eQ(f)\bar{f}\gamma^\mu f A_\mu-\fr{g_L}{2c_W}\bar{f}\gamma^\mu [g_V^{Z}(f)-g_A^{Z}(f)\gamma_5]f Z_\mu\crn
&&-\fr{g_L}{2c_W}\bar{f}\gamma^\mu [g_V^{\mathcal{Z}_{1}}(f)-g_A^{\mathcal{Z}_{1}}(f)\gamma_5]f \mathcal{Z}_{1\mu}-\fr{g_L}{2c_W}\bar{f}\gamma^\mu [g_V^{\mathcal{Z}_{1}^\prime}(f)-g_A^{\mathcal{Z}_{1}^\prime}(f)\gamma_5]f \mathcal{Z}_{1\mu}^\prime,
\label{hu}\eea where $f$ indicates to every fermion of the model and note that $Q(f_L)=Q(f_R)=Q(f)$ as well as $e=g_L s_W$. The vector and axial-vector couplings $g^{Z,\mathcal{Z}_1,\mathcal{Z}'_1}_{V,A}(f)$ can be directly obtained from the corresponding chiral couplings in the expressions of $P^{NC}_{L,R\mu}$ above to yield,     
\bea 
g^{Z}_V(f) &=& T_{3L}(f_L)-2s^2_W Q(f),\hs g^{Z}_A(f) = T_{3L}(f_L),\crn
g^{\mathcal{Z}_{1}}_V(f) &=& \frac{t_W^2 (\beta t_Xc_\epsilon c_W +s_\epsilon s_W) [T_{3L}(f_L)-2Q(f)]+s_W(t_R^2s_\epsilon+t_Xt_W\beta c_\epsilon)T_{3R}(f_R)}{t_X^{-1}t_W^2\sqrt{t_R^2+ t_X^2\beta^2}}\crn
&&+c_\epsilon c_W\sqrt{t_R^2+t_X^2\beta^2} T_{8R}(f_R),\crn
g^{\mathcal{Z}_{1}}_A(f) &=&\frac{t_W^2 (\beta t_Xc_\epsilon c_W +s_\epsilon s_W)T_{3L}(f_L)-s_W(t_R^2s_\epsilon +t_Wt_X\beta c_\epsilon)T_{3R}(f_R)}{t_X^{-1}t_W^2\sqrt{t_R^2+ t_X^2\beta^2}}\crn
&&-c_\epsilon c_W\sqrt{t_R^2+t_X^2\beta^2} T_{8R}(f_R),\crn
g^{\mathcal{Z}_{1}^\prime}_{V,A}&=&g^{\mathcal{Z}_{1}}_{V,A}(c_\epsilon \rightarrow s_\epsilon,\ s_\epsilon \rightarrow -c_\epsilon).
\eea 

The first term in (\ref{hu}) yields electromagnetic interactions, as usual. The second term in (\ref{hu}) determines the neutral current coupled to $Z$ boson, which is consistent with the standard model. Note that the couplings of $\mathcal{Z}_{1}^\prime$ can be obtained from those of $\mathcal{Z}_{1}$ by replacing $c_\epsilon\rightarrow s_\epsilon,\ s_\epsilon \rightarrow -c_\epsilon$, and vice versa. All the vector and axial-vector couplings of $Z,\ \mathcal{Z}_1,\ \mathcal{Z}'_1$ with fermions are explicitly calculated as collected in Appendix \ref{gvaf}. 

 \subsection{Scalar--gauge boson interactions}
 
The interactions of gauge bosons with scalars arise from \eqref{gma1}. First note that there is no strong interaction for the scalars since they are colorless. Next, expand the scalar fields around their VEVs as in \eqref{kthiggs} and \eqref{higgs}. Substituting the physical scalar states from \eqref{lhh} and the physical gauge states from \eqref{gaugec1}, \eqref{gaugec2}, and \eqref{gaugen} into the mentioned Lagrangian, we get desirable interactions according to the vertex types between a gauge boson and two scalars, a scalar and two gauge bosons, and two scalars and two gauge bosons in the model\footnote{See Appendix B in the first version of the arXiv posting of this article, arXiv:1609.03444v1 [hep-ph], for the detailed derivations of the Feynman rules corresponding to the various vertices between the scalar and gauge fields and the associated couplings, which were appropriately listed from Table IV to Table XXIII.}. Consequently, all the standard model interactions between the Higgs boson and the gauge fields are consistently recovered at the leading order.

\section{\label{flavorsec} New physics effects and constraints}

In Ref. \cite{3L3R}, the 750 GeV diphoton excess reported by the ATLAS and CMS experiments was studied, and the ATLAS diboson anomalies were briefly discussed too. Since these signals disappeared in the early search results of the LHC run II, the new physics scales should be high enough and their masses should be correspondingly large, above several TeVs, to escape the detections. Of course, the electric charge parameter $q$ could be kept compatible to the usual ones. Retaining these conditions, in this work we will pay attention to alternative, interesting new-physics features that include the mixing effects in gauge and scalar sectors as well as the tree-level FCNCs.

\subsection{$\rho$ and mixing parameters}

The new physics contribution to the $\rho$-parameter starts from the tree-level due to both mixings of the standard model $Z$ and $W$ bosons with new gauge bosons. It is evaluated as 
\bea \Delta \rho &\equiv& \rho-1=\fr{m^2_{W_1}}{c^2_W m^2_{\mathcal{Z}}}-1\crn
&\simeq& \epsilon_2 \fr{t^2_R c^3_W \kappa}{[t^2_R+t^2_X(1+\beta^2)]^{3/2}}+\epsilon_1\fr{(v^2-u^2)c^3_W\kappa}{\sqrt{3}(u^2+v^2)[t^2_R+t^2_X (1+\beta^2)]}-\fr{2u^2v^2}{(u^2+v^2)\La^2},\eea which is suppressed due to $u,v\ll w,\La$. 

\begin{figure}[H]
 \centering
  \includegraphics[scale=0.6]{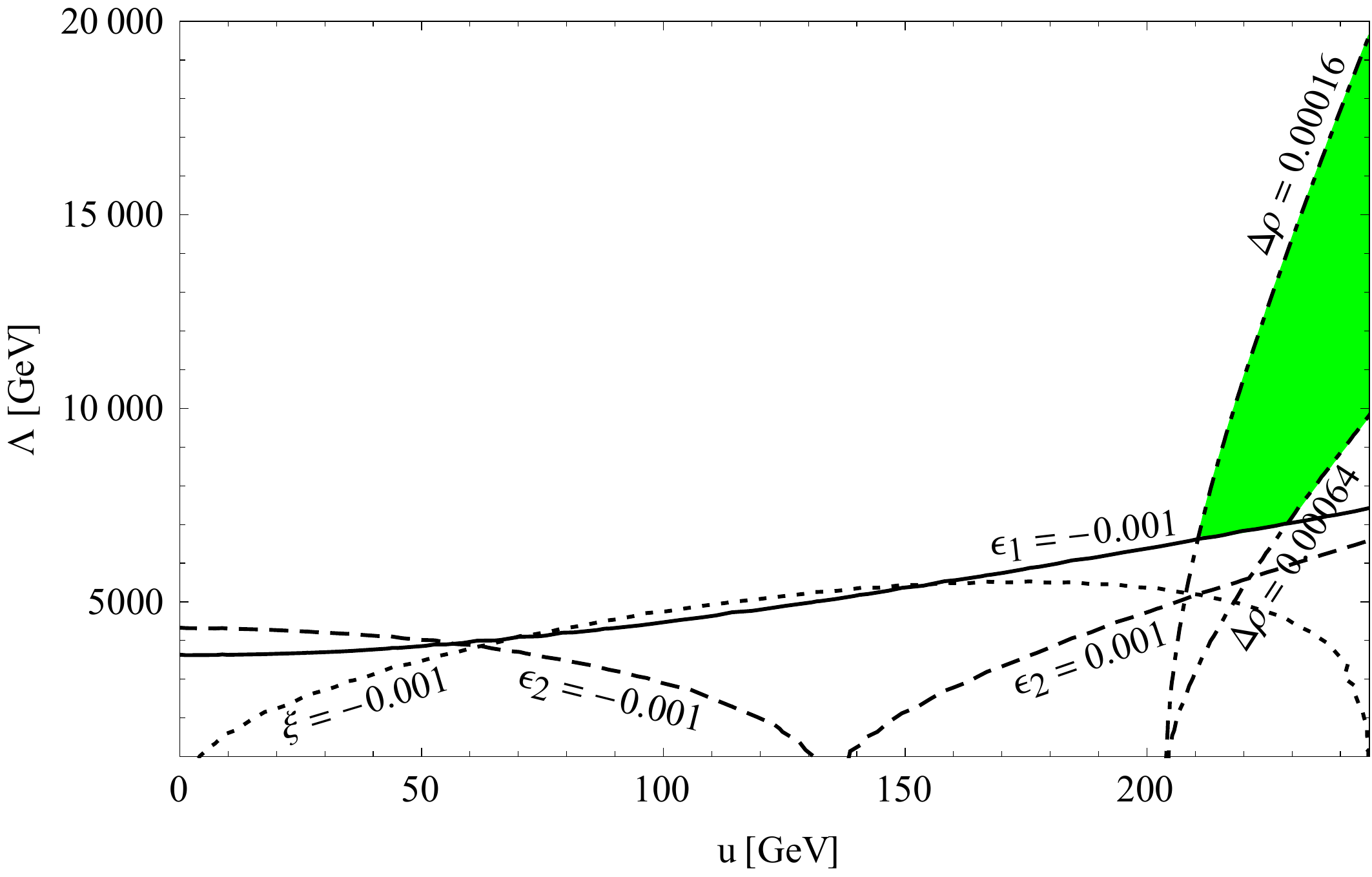}
   \caption{\label{rhomixa}The viable new physics regime (green) as constrained by $0.00016<\Delta\rho<0.00064$, $\xi=\epsilon_1=\epsilon_2=\pm 10^{-3}$ for the case $\beta=-1/\sqrt{3}$.}
   \end{figure}

\begin{figure}[H]
 \centering
  \includegraphics[scale=0.6]{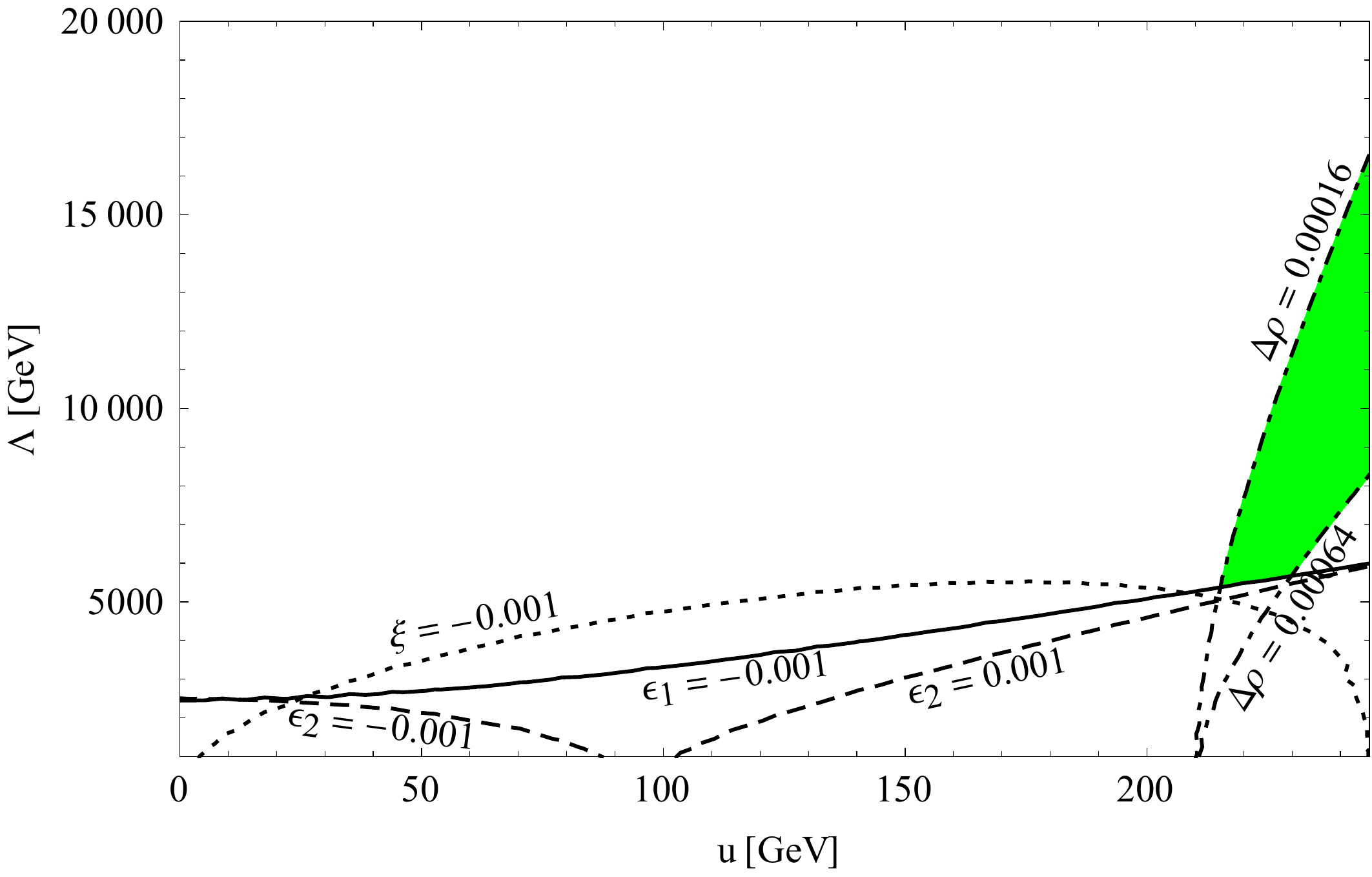}
   \caption{\label{rhomixk}The viable new physics regime (green) as constrained by $0.00016<\Delta\rho<0.00064$, $\xi=\epsilon_1=\epsilon_2=\pm 10^{-3}$ for the case $\beta=0$.}
   \end{figure}

\begin{figure}[H]
 \centering
  \includegraphics[scale=0.6]{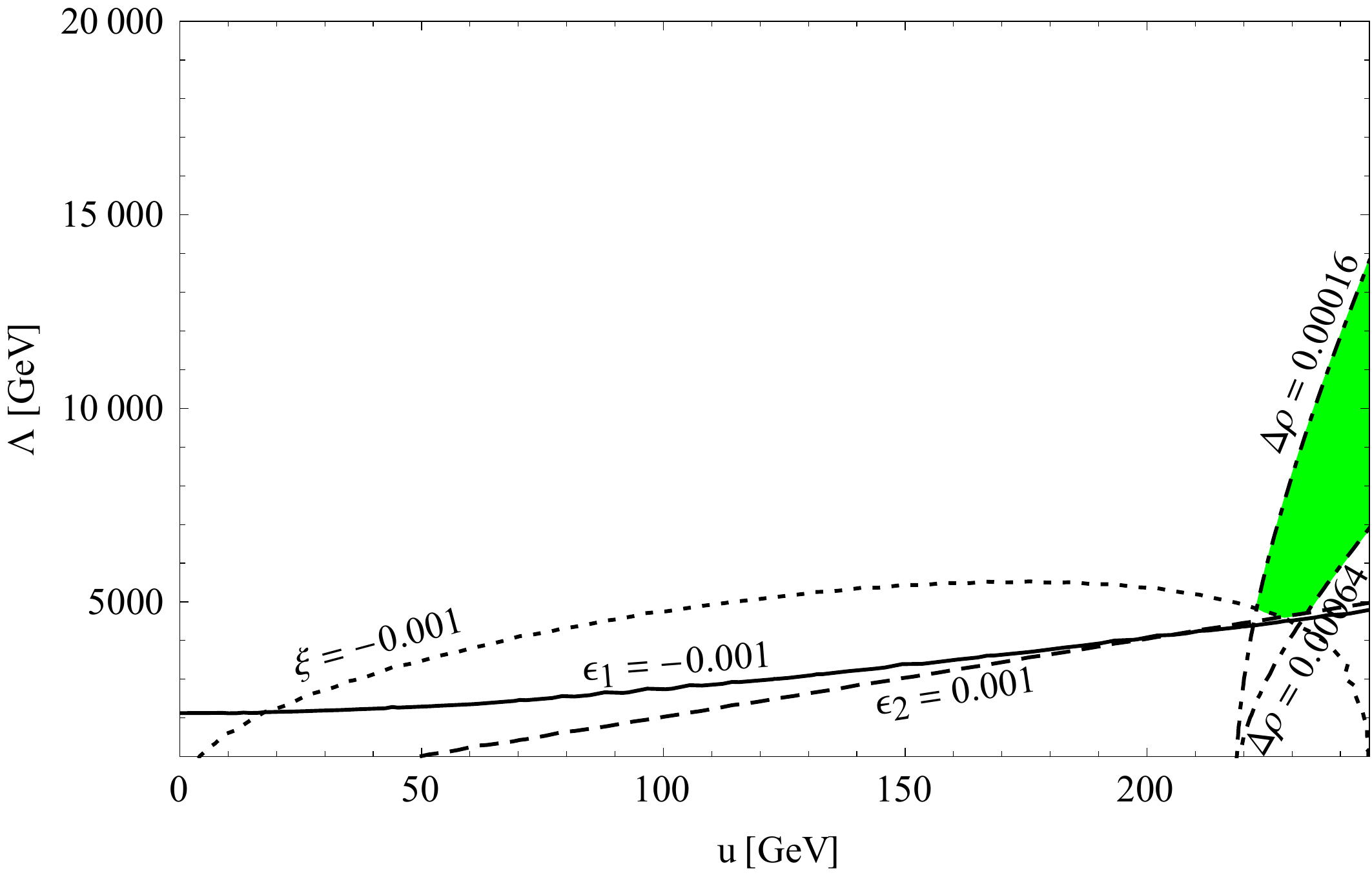}
   \caption{\label{rhomixd}The viable new physics regime (green) as constrained by $0.00016<\Delta\rho<0.00064$, $\xi=\epsilon_1=\epsilon_2=\pm 10^{-3}$ for the case $\beta=1/\sqrt{3}$.}
   \end{figure}

The $W$ mass implies $u^2+v^2=(246\ \mathrm{GeV})^2$. Further, we take $t_R=g_R/g_L=1$ and thus $t_X=s_W/\sqrt{1-(2+\beta^2)s^2_W}$. Note also that $|\beta|<\sqrt{1/s^2_W-2}\simeq 1.5261$, provided $s^2_W\simeq 0.231$. From the global fit, the $\rho$-parameter is constrained by $\rho=1.0004\pm0.00024$, which is 1.7$\sigma$ deviating from the standard model prediction, $\rho=1$ \cite{data}. If the data imply a potential new physics, it sets the corresponding new physics scale via $0.00016 < \Delta \rho < 0.00064$ at 95\% CL. Otherwise, when the measured central value is due to statistic errors, it induces a lower bound on the new physics scale of interested model via $\Delta\rho < 0.00064$ at 95\% CL. For this case, note that an upper bound on the new physics scale is not presented. In the following, we take the first interpretation into account. We make a contour (long and short dashed line) for $\Delta \rho$ as a function of $\La=w=1$-20 TeV and $u=0$-246 GeV for three cases $\beta=-1/\sqrt{3}$, 0, and $1/\sqrt{3}$ as in Figs. \ref{rhomixa}, \ref{rhomixk}, and \ref{rhomixd}, respectively. The available parameter space is bounded by both the lines of respective $\Delta \rho$ values. 

The mixing of $W,\ Z$ bosons with the new gauge bosons also modifies the well-measured couplings of $W,Z$ with fermions. This new physics effect is safe if one imposes the mixing parameters $\xi$, $\epsilon_{1,2}$ in $10^{-3}$ \cite{data}. (Recall that these parameters are very suppressed due to the condition, $u,v\ll w,\La$, too). In Figs. \ref{rhomixa}, \ref{rhomixk}, and \ref{rhomixd}, we make contours (solid line for $\epsilon_1$, dashed line for $\epsilon_2$, and short dashed line for $\xi$) for $|\xi|=|\epsilon_{1,2}|=10^{-3}$ in terms of ($\La=w,u$), using the above inputs. The available parameter space lies above these three lines. 

Combining all the constraints, the new physics regime is those as green-colored (grey in print) as also included in Figs \ref{rhomixa}, \ref{rhomixk}, and \ref{rhomixd} for the three cases of $\beta$ aforementioned. Consequently, $\La$ (thus $w=\La$) is bounded by $4.6\ \mathrm{TeV}<\La<13.7\ \mathrm{TeV}, 5.5\ \mathrm{TeV}<\La< 16.3\ \mathrm{TeV}$, and $6.6\ \mathrm{TeV}<\La<19.4\ \mathrm{TeV}$ for $\beta=1/\sqrt{3},\ 0$, and $-1/\sqrt{3}$, respectively. The weak scale regime for $u$ (thus $v=\sqrt{(246\ \mathrm{GeV})^2-u^2}$ is followed) is narrow, as limited by $u<246$ GeV, and $u>222.3$, 215, and 210.4 GeV corresponding to the $\beta$ values as mentioned.       

Particularly, the bounds for the 3-2-3-1 scales ($w,\La$) that come from the $\rho$-parameter depend significantly on the weak scale, $u$, and that they may even relax to zero for the certain values of $u$ with regard to the corresponding $\beta$ values. But, this is not the complete story that we hope to close the 3-2-3-1 symmetry at the weak scale, like the 3-3-1 models investigated in \cite{close331}. As a matter of the fact, although $\Delta \rho$ is proportional to $\epsilon_{1,2}$ and $\xi\sim uv/\La^2$ (with the finite coefficients) according to the mixing of $Z$ with ($Z_R, Z'_R$) and the mixing of $W_L$ with $W_R$, the new physics is not decoupled from the standard model when $w,\La$ tend to the weak scale or even to zero. Indeed, the mixing effects and thus the $W,Z$-coupling corrections diverge when $(w,\La)\rightarrow 0$ along the $\Delta \rho$ bounds, as possibly seen from the corresponding figures for $\epsilon_{1,2}$ and $\xi$ (even though these mixing effects cancel out in the $\Delta\rho$ expression). Apparently, this property also emerges at loop levels because the good custodial symmetry $SU(2)_{L+R}$, if imposed, only protects $\rho$ from the large contributions due to the effective cancelations, but it does not preserve any individual mixing effect from the divergences. The above judgement is also valid for arbitrary $\beta$ and $w$-$\La$ relation. Therefore, closing new gauge symmetries at the weak scale as observed in the 3-3-1 models would be lost due to the contribution of the several new gauge bosons (not one) to the $\rho$-parameter.

It is easily checked that when $(w,\La)$ go infinity, we have $\rho\rightarrow 1$ since $\epsilon_{1,2}$ and $\xi$ as well as the loop effects of the new gauge boson and scalar doublets are suppressed by $(u^2,v^2)/(w^2,\La^2)$ as the mass-squared splittings of the doublet components are. The standard model like fields and masses as well as couplings are restored to the standard model. Therefore, the 3-2-3-1 model has a decoupling limit at a high scale for $(w,\La)$, not at the small scale as analysized above. 

The running of the gauge couplings and scalar self-couplings along with the energy scale may potentially present a upper bound on the 3-2-3-1 breaking scales, such as the Landau pole at which some gauge coupling becomes infinity or the metastable scale at which some scalar self-coupling becomes negative. This model predicts the Weinberg angle as \be s^2_W=t^2_R t^2_X/[t^2_R+t^2_X(1+\beta^2+t^2_R)]<t_R^2/(1+\beta^2+t_R^2),\label{dchcx}\ee where $g_{L,R}$ small change as $t_R=g_R/g_L$ does, while $g_X$ and thus $t_X=g_X/g_L$ significantly raises, with the increasing of the energy scale. Therefore, the model encounters a Landau pole ($M$) at which $s^2_W(M)=t_R^2/(1+\beta^2+t_R^2)<1$ or $g_X(M)=\infty$. Of course, the consistent condition of the theory is $w,\La<M$. For simplicity, we assume $t_R=1$, thus $s^2_W(M)=1/(2+\beta^2)$, without much loss of generality ($t_R=1$ is possibly protected by a minimal left-right symmetry, as also used around). The condition (\ref{dchcx}), $s^2_W(M)>s^2_W\simeq 0.231$ at the weak scale, implies $|\beta|<1.5261$, which translates to $-1.821<q<0.821$, which is very constrained, but spans every elementary charge as observed. Although the $q$ charge is arbitrary in its range, the model predicts only integer charges for $q=0,-1$. When $q$ coincides its bounds $q=-1.821$ or $0.821$, the Landau pole lies at the weak scale $M\sim v_{\mathrm{weak}}$. Since the new physics is not decoupled as shown, the model in this case is inconsistent. For the half integer charges $q=0.5,-1.5$ which are near the corresponding bounds, the Landau pole is lifted much $M\sim 10$ TeV, by which the new physics can be interestingly explored at the current colliders. For $q=0,-1/2,-1$, which are being taken throughout the text, the Landau pole may be higher than the Planck scale. The above conclusions are analogous to the case of the 3-3-1 models as studied in \cite{landau}.

\subsection{FCNC}

As described above, after the spontaneous symmetry breaking, the Yukawa Largangian yields the masses for the fermions. Therefore, we will extract the quark mass terms from (\ref{231}). The exotic quarks get large masses at the $w$ scale,
 \bea
 \mathcal{L}^{J}_{\mathrm{mass}} = \bar{J}_{3L} \frac{h_{33}^J w}{\sqrt{2}}J_{3R} +\bar{J}_{\al L} \frac{h_{\al \beta}^J w}{\sqrt{2}}J_{\beta R}+ H.c.,
 \eea
 which are physical and decoupled (i.e., do not mix with the ordinary quarks and can be integrated out). Whilst, the ordinary quarks mix by themselves with a mass Lagrangian given by
 \bea
 \mathcal{L}^{u,d}_{\mathrm{mass}} =  -\sum_{a,b} \bar{u}_{aL} \mathcal{M}^U_{ab} u_{bR} -  \sum_{a,b} \bar{d}_{aL} \mathcal{M}^D_{ab} d_{bR} + H.c.,
 \eea
 where
 \bea
 \mathcal{M}^U &=& \{\mathcal{M}^U_{ab} \} =-\frac{1}{\sqrt{2}} \left( \begin{array}{ccc} h_{11}^q v& h_{12}^q v& h^q_{13}u \\
 h_{21}^q v & h_{22}^q v & h_{23}^q u \\ h_{31}^q v & h_{32}^q v & h_{33}^q u \end{array} \right),\\
 \mathcal{M}^D &=& \{\mathcal{M}^D_{ab} \} =-\frac{1}{\sqrt{2}} \left( \begin{array}{ccc} h_{11}^q u& h_{12}^q u& h^q_{13}v \\
 h_{21}^q u & h_{22}^q u & h_{23}^q v \\ h_{31}^q u & h_{32}^q u & h_{33}^q v \end{array} \right),
 \eea
 which are generally complex-valued matrices and correlated due to $u\neq v$.

 By applying bi-unitary transformations, we can diagonalize the mass matrices, $\mathcal{M}^U$ and $\mathcal{M}^D$, separately, such that
 \bea
 V_{dL}^\dag \mathcal{M}^D V_{dR}= M^D, \hs V^\dag_{uL} \mathcal{M}^U V_{uR} = M^U,
 \eea
 where $M^U, M^D$ are diagonal matrices and $V_{uL,R}, V_{dL,R}$ are unitary matrices. The mass eigenstates and gauge states
 are related by
 \bea
 d_{L, R} =V_{d L,R} d_{L, R}^\prime,  \hs  u_{L, R} =V_{u L, R} u_{L, R}^\prime,
 \eea
 where we use the notations, the gauge states for up-quarks $u=(u_1, u_2, u_3)^T$, for down-quarks $d =(d_1, d_2, d_3)^T$, and the mass eigenstates $u^\prime = (u,c,t)^T$, $d^\prime =(d, s, b)^T$. The CKM matrix is defined as $V_{\mathrm{CKM}}= V_{uL}^\dag V_{dL}$. Note also that although the up and down quark mass matrices differ by only a relation, $u\neq v$, the realistic masses for the quarks can be achieved by choosing appropriate parameters. Even if $h^q_{ab}$ is flavor-diagonal, we need only $u\gg v$ and $h^q_{33}\gg h^q_{11,22}$. In this case, there are only two unsuitably-small masses corresponding to $u,c$ as well as the small quark mixing angles, which can be radiatively induced.

We would like to emphasize that two of three right-handed quark multiplets transforming differently from the remainder under $SU(3)_R$. This causes the FCNCs at the tree level for the ordinary quarks due to the followings, \ben \item {\it Nonuniversal gauge ($Z'_R$) couplings}: The flavors of ordinary quarks such as $\{u_a\}$ and $\{d_a\}$ differ in $T_{8R}$ as well as $X$ charges (note that all the lepton flavors as $\{\nu_a\}$, $\{e_a\}$, $\{E_a\}$ and exotic quark flavors $\{J_\al\}$ do not have this property since the corresponding left or right flavors in each group are identical under every neutral gauge charge; also, there is no flavor changing associated with $Q$, $T_{3L,R}$ since each of them couples universally to every left or right flavor groups aformentioned). Since $X$ is related to $T_{8R}$, the FCNCs are mediated by only the extra neutral gauge boson,
$Z^\prime_R$, which couples to $T_{8R}$. \item {\it Nonuniversal Higgs ($H_2$) couplings}: Although the Higgs doublets are unified in $S$, the FCNCs associated the ordinary quarks arise due to the nonuniversal arrangement of quark generations under the gauge symmetry. This can be seen from the Yukawa interactions for $S$ and quarks. Similarly to the previous case, there is no flavor changing associated with the other fermions as well as other neutral scalars. A combination of $S_{11}$ and $S_{22}$ is just the standard model Higgs boson, $H_1$, which conserves flavors since its Yuakwa couplings are proportional to the corresponding quark mass matrices. However, the new Higgs state, $H_2$, which is directly orthogonal to $H_1$ changes flavors.\een

First, let us consider the FCNCs induced from quark and scalar interactions. The same Yukawa terms in (\ref{231}) that yield the quark masses also bring FCNCs into the up and down quark sectors,
 \bea
 \mathcal{L}^{u,d}_{\mathrm{int}} &=&  h_{a3}^q\bar{d}_{aL} S_{22}^0 d_{3R}+ h_{a \beta}^q \bar{d}_{aL} S_{11}^0 d_{\beta R}+ h_{a3}^q\bar{u}_{aL} S_{11}^0 u_{3R}+ h_{a \beta}^q \bar{u}_{aL} S_{22}^0 u_{\beta R}+
H.c.
  \crn & =&
  h_{a3}^q\bar{d}_{aL} \frac{uH_2+vH_1}{\sqrt{2(u^2+v^2)}} d_{3R}+ h_{a \beta}^q \bar{d}_{aL}
  \frac{u H_1-vH_2}{\sqrt{2(u^2+v^2)}} d_{\beta R}
\crn  && +
  h_{a3}^q\bar{u}_{aL} \frac{u H_1-vH_2}{\sqrt{2(u^2+v^2)}} u_{3R}+ h_{a \beta}^q \bar{u}_{aL}
 \frac{uH_2+vH_1}{\sqrt{2(u^2+v^2)}} u_{\beta R}+ H.c.\crn
  &=&-\bar{d}^\prime_L \frac{M^D}{\sqrt{(u^2+v^2)}}  d^\prime_R H_1
  +\frac{v}{u} \bar{d}^\prime_L \frac{M^D}{\sqrt{(u^2+v^2)}}  d^\prime_R H_2 \crn
  &&-\bar{u}^\prime_L \frac{M^U}{\sqrt{(u^2+v^2)}}  u^\prime_R H_1
  -\frac{u}{v} \bar{u}^\prime_L \frac{M^U}{\sqrt{(u^2+v^2)}}  u^\prime_R H_2
\crn &&-\frac{\sqrt{u^2+v^2}}{u^2}\bar{d}^\prime_{iL} (V_{dL}^\dagger
  V_{uL})_{ik}(M^U)_{km}(V_{uR}^*)_{3m}(V_{dR})_{3j}d^\prime_{jR}H_2 \crn
 &&+\frac{\sqrt{u^2+v^2}}{v^2}\bar{u}^\prime_{iL} (V_{uL}^\dagger
  V_{dL})_{ik}(M^D)_{km}(V_{dR}^*)_{3m}(V_{uR})_{3j}u^\prime_{jR}H_2 + H.c.
 \label{FCNC1}\eea

 We see that the Higgs boson, $H_1$, couples to quarks, even charged leptons, similarly to those in the standard model, which is a feature vadidating this model \cite{Higgsd}. $H_2$ is a new heavy Higgs boson, which changes quark flavors, as desirable, presented by the non-zero off-diagonal elements ($i\neq j$) in the last two terms of (\ref{FCNC1}). Therefore, the tree level FCNC processes might appear due to the contribution of $H_2$ as mediators. Conventionally, we rewrite the relevant couplings as follows
 \bea
\mathcal{L}_{\mathrm{FCNC}}^{H_2} = \bar{d}^\prime_{iL} \Gamma^d_{ij}d^\prime_{jR}H_2
+\bar{u}^\prime_{iL}\Gamma^u_{ij} u^\prime_{jR}H_2 + H.c.,
 \eea
where \bea \Gamma^d_{ij} &=& -\frac{\sqrt{u^2+v^2}}{u^2} (V_{dL}^\dagger
  V_{uL})_{ik}(M^U)_{km}(V_{uR}^*)_{3m}(V_{dR})_{3j}, \crn
 \Gamma^u_{ij}&=&\frac{\sqrt{u^2+v^2}}{v^2} (V_{uL}^\dagger
  V_{dL})_{ik}(M^D)_{km}(V_{dR}^*)_{3m}(V_{uR})_{3j}.
   \eea

Second, we consider the FCNCs due to the fermion and gauge boson interactions. As mentioned, the FCNCs associated with $Z'_R$ are due to the third generation of quarks transforms differently from the first two under the gauge symmetry. Here, the FCNCs occur in the right-handed quark sector and with the
gauge bosons, $A_{8R}$ and $B$, which couple to $T_{8R}$ and $X$, respectively. Since $X=Q - T_{3L}-T_{3R}- \beta T_{8R}$, the source for the FCNCs is only $T_{8R}$. Indeed, considering the interacting Lagrangian of neutral gauge bosons with fermions and using the expression for $X$, we come to the relevant interaction,
 \bea
\mathcal{L}_8 &=& - \sum_{a=1}^3 \bar{Q}_{aR} \ga^\mu T_{8R} Q_{aR} (g_R A_{8R\mu}-\beta g_X B_\mu) = - g_L \sqrt{t_R^2+\beta^2 t_X^2} \sum_{a=1}^3 \bar{Q}_{aR} \ga^\mu T_{8R} Q_{aR} Z'_{R \mu} \crn
&\supset&  -g_L \sqrt{t_R^2+\beta^2 t_X^2}(\bar{u}_R \ga^\mu T_u u_R+ \bar{d}_R \ga^\mu T_d d_R)Z'_{R \mu} \crn
&=&
-g_L \sqrt{t_R^2+\beta^2 t_X^2}\left(\bar{u}^\prime_R \ga^\mu (V^\dag_{uR} T_u V_{uR})u^\prime_R+\bar{d^\prime}_R \ga^\mu (V^\dag_{dR} T_d V_{dR})d^\prime_R \right) Z'_{R \mu},
 \eea
 where $T_u =T_d =\frac{1}{2 \sqrt{3}}\mathrm{diag}(-1,-1,1)$ includes $T_{8R}$ values of up or down quark flavors.
 The tree-level FCNC associated with the field $Z'_{R}$ is obtained by
 \bea
 \mathcal{L}^{Z'_{R}}_{\mathrm{FCNC}} = -\Theta^{Z^\prime_R}_{ij}\bar{q}^\prime_{iR} \ga^\mu q^\prime_{jR} Z'_{R \mu}
 \eea
 with $i \neq j$, where $q^\prime$ is denoted as either $u^\prime$ or $d^\prime$, and
 $\Theta^{Z^\prime_R}_{ij}$ is defined as
\bea \Theta^{Z^\prime_R}_{ij}=\frac{g_L}{\sqrt{3}} \sqrt{t_R^2+\beta^2
t_X^2}(V^*_{qR})_{3i}(V_{qR})_{3j}. \eea

In the following, we will calculate the contribution of the new physics to
the meson mixing systems as mediated by the neutral scalar $H_2$ and neutral gauge boson $Z'_{R}$.
For the case of the $K^0$-$\bar{K}^0$ mixing, the relevant effective Lagrangian is given after integrating out $H_2$ and $Z'_R$,
 \bea
 \mathcal{L}_{{\mathrm{effective}}}^{\Delta S=2} &=& -\frac{(\Theta^{Z^\prime_R}_{12})^2}{m^2_{Z'_{R}}}(\bar{d}_R \ga^\mu
 s_R)^2+\frac{(\Gamma_{12}^d)^2}{m_{H_2}^2}(\bar{d}_L s_R)^2+\frac{(\Gamma_{21}^{d*})^2}{m_{H_2}^2}(\bar{d}_R s_L)^2
 \crn &&+ \frac{\Gamma_{21}^{d*}\Gamma_{12}^d}{m_{H_2}^2}(\bar{d}_L s_R)(\bar{d}_R s_L)+
 \frac{\Gamma_{21}^{d*}\Gamma_{12}^d}{m_{H_2}^2}(\bar{d}_R s_L)(\bar{d}_L s_R).
\label{kk1} \eea This yields the contribution to
the $K^0$-$\bar{K}^0$ mixing parameter or mass difference $\Delta m_K$ as  \bea
\Delta m_K = 2\mathrm{Re}\langle \bar{K}^0\vert
- \mathcal{L}_{\mathrm{eff}}^{\Delta S =2} \vert K^0 \rangle.
\label{kk4}\eea Using the matrix elements
\cite{matrixelements} \bea \langle \bar{K}^0\vert (\bar{d}_R
\ga^\mu
 s_R)^2 \vert K^0 \rangle &=&\langle \bar{K}^0\vert(\bar{d}_L\ga^\mu
 s_L)^2\vert K^0 \rangle=\frac{1}{3}m_K f_K^2, \crn
  \langle \bar{K}^0\vert(\bar{d}_L
 s_R)^2 \vert K^0 \rangle &=& \langle \bar{K}^0\vert (\bar{d}_R
 s_L)^2 \vert K^0 \rangle= -\frac{5}{24}\left(\frac{m_K}{m_s+m_d} \right)^2 m_K f_K^2, \crn
 \langle \bar{K}^0\vert(\bar{d}_L
 s_R)(\bar{d}_R
 s_L) \vert K^0 \rangle &=& \langle \bar{K}^0\vert (\bar{d}_R
 s_L)(\bar{d}_L
 s_R)\vert K^0 \rangle= \left [\frac{1}{24}+\frac{1}{4}\left(\frac{m_K}{m_s+m_d} \right)^2 \right ] m_K f_K^2,\nn
\eea the $K^0$-$\bar{K}^0$ mixing parameter $\Delta m_K$ is
obtained by \bea \Delta m_K &=&\mathrm{Re}\left\{\frac{2}{3}
\frac{(\Theta^{Z^\prime_R}_{12})^2}{m^2_{Z^\prime_{R}}}+\frac{5}{12}\left(\frac{(\Gamma_{21}^{d*})^2}{m_{H_2}^2}+
\frac{(\Gamma_{12}^{d})^2}{m_{H_2}^2}\right)\left(\frac{m_K}{m_s+m_d}
\right)^2-\frac{\Gamma_{21}^{d*}\Gamma_{12}^d}{m_{H_2}^2}\left[
\frac{1}{6}+\left(\frac{m_K}{m_s+m_d} \right)^2\right] \right\}
\crn &&\times   m_K f_K^2.\eea 

Similarly, we obtain
$B^0_{d,s}$-$\bar{B}^0_{d,s}$ mixing parameters, $\Delta m_{B_d}$ and $\Delta m_{B_s}$, as \bea \Delta m_{B_d}
&=&\mathrm{Re}\left\{\frac{2}{3}
\frac{(\Theta^{Z^\prime_R}_{13})^2}{m^2_{Z^\prime_{R}}}+\frac{5}{12}\left(\frac{(\Gamma_{31}^{d*})^2}{m_{H_2}^2}+
\frac{(\Gamma_{13}^{d})^2}{m_{H_2}^2}\right)\left(\frac{m_{B_d}}{m_b+m_d}
\right)^2-\frac{\Gamma_{31}^{d*}\Gamma_{13}^d}{m_{H_2}^2}\left[
\frac{1}{6}+\left(\frac{m_{B_d}}{m_b+m_d} \right)^2\right] \right\}
\crn &&\times   m_{B_d} f_{B_d}^2,\\
\Delta m_{B_s}
&=&\mathrm{Re}\left\{\frac{2}{3}
\frac{(\Theta^{Z^\prime_R}_{23})^2}{m^2_{Z^\prime_{R}}}+\frac{5}{12}\left(\frac{(\Gamma_{32}^{d*})^2}{m_{H_2}^2}+
\frac{(\Gamma_{23}^{d})^2}{m_{H_2}^2}\right)\left(\frac{m_{B_s}}{m_b+m_s}
\right)^2-\frac{\Gamma_{32}^{d*}\Gamma_{23}^d}{m_{H_2}^2}\left[
\frac{1}{6}+\left(\frac{m_{B_s}}{m_b+m_s} \right)^2\right] \right\}
\crn &&\times   m_{B_s} f_{B_s}^2.\eea

Let us numerically study the mixing parameters, $\Delta m_K$ and $\Delta m_{B_{d,s}}$, by
using the following input parameters (mass parameters are measured in MeV) \cite{paras,data,data1}: \bea && m_d = 4.73 ,
\hs m_s= 93.4, \hs m_b =4190, \hs m_t =173 \times 10^3,\hs f_K =156.1, \crn
&& m_K=497.614,\hs f_{B_d}=188,\hs m_{B_d}=5279.5,\hs f_{B_s}=225,\hs m_{B_s}=5366.3,\crn
&& (V_{\mathrm{CKM}})_{31}=0.00886,\hs (V_{\mathrm{CKM}})_{32}=0.0405,\hs (V_{\mathrm{CKM}})_{33}=0.99914. \eea Referring to the above results for the weak scales, we take $u=230$ GeV, and thus $v$ is followed from $u^2+v^2=(246\ \mathrm{GeV})^2$. Also, $t_R=1$, $t_X=s_W/\sqrt{1-(2+\beta^2)s^2_W}$, and $s^2_W=0.231$ as given before are used. For the above $\beta$ values, i.e. $\beta=0,\pm 1/\sqrt{3}$, $t_X$ and $\Theta^{Z'_R}_{ij}$ slightly change. So, we can take $|\beta|=1/\sqrt{3}$ for further calculations. We have $g_L=\sqrt{4\pi \al/s^2_W}$, with $\al=1/128$. For the right-handed quark mixing matrices, $V_{qR}\ (q=u,d)$, the elements that enter the meson mass differences, $\Delta m_{K,B_d,B_s}$, are $(V_{uR})_{33}$, $(V_{dR})_{31}$, $(V_{dR})_{32}$, and $(V_{dR})_{33}$. Since $\Delta m_{K,B_d,B_s}$ depend symmetrically on $(V_{dR})_{31}$ and $(V_{dR})_{32}$, one can assume $(V_{dR})_{31} = (V_{dR})_{32}\equiv V_{dR}$ without loss of generality. Thus, $(V_{dR})^2_{33}=1-2V^2_{dR}$ due to the unitarity. We also label $(V_{uR})_{33}\equiv V_{uR}$ for simplicity. As seen, the contributions of $H_2$ and $Z'_R$ are compatible. So, let us simply take $m_{H_2}=m_{Z'_R}\equiv M$, which are commonly called the new physics scale as entering the flavor changing processes.  

The standard model contributions to the meson mass differences are given by \cite{dmkbdsm}, 
\be (\Delta m_K)_{\mathrm{SM}}=0.467\times 10^{-2}/ps,\hs (\Delta m_{B_d})_{\mathrm{SM}}=0.528/ps,\hs (\Delta m_{B_s})_{\mathrm{SM}}=18.3/ps.\ee
Whereas, the experimental values are \cite{dmkbdsm} 
\be (\Delta m_K)_{\mathrm{Exp}}=0.5292\times 10^{-2}/ps,\hs (\Delta m_{B_d})_{\mathrm{Exp}}=0.5055/ps,\hs (\Delta m_{B_s})_{\mathrm{Exp}}=17.757/ps. \ee
Note that the meson mass differences of the considering model are given by \be (\Delta m_{K,B_d,B_s})_{\mathrm{tot}}=(\Delta m_{K,B_d,B_s})_{\mathrm{SM}}+\Delta m_{K,B_d,B_s},\ee where the last terms are due to the new physics contributions, which have been obtained above. These total contributions will be compared with the experimental values. We require the theory to produce the data for the kaon mass difference within 30\% due to the potential long-range uncertainties, whereas it is within 5\% for the B-meson mass differences, namely \bea 
&& 0.37044\times 10^{-2}/ps<(\Delta m_K)_{\mathrm{tot}}<0.68796\times 10^{-2}/ps,\\ 
&& 0.480225/ps<(\Delta m_{B_d})_{\mathrm{tot}}<0.530775/ps,\\
&& 16.8692/ps<(\Delta m_{B_s})_{\mathrm{tot}}<18.6449/ps. \eea   

\begin{figure}[H]
 \centering
 \begin{tabular}{cc}
  \includegraphics[width=8cm]{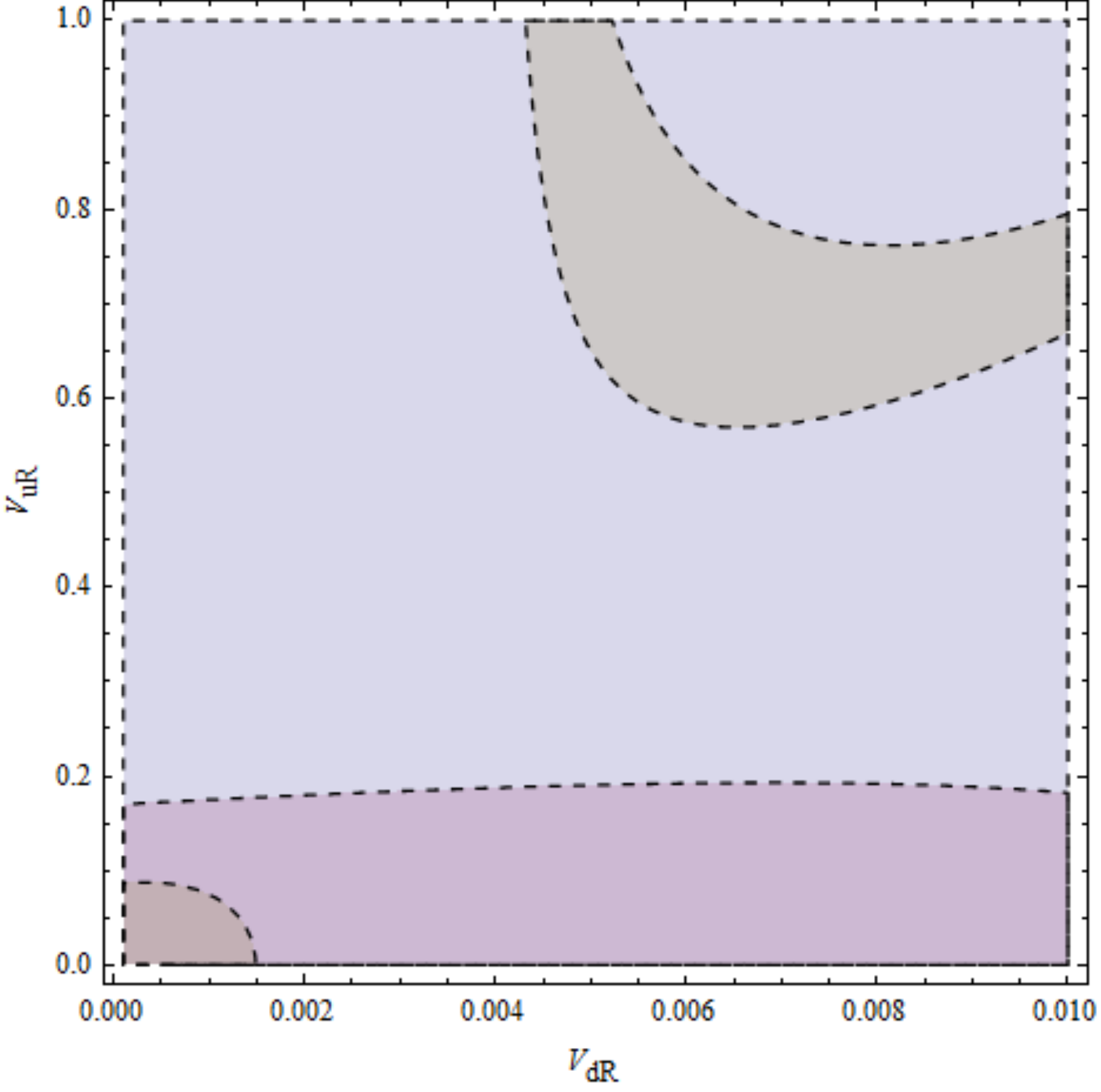} &
  \includegraphics[width=8cm]{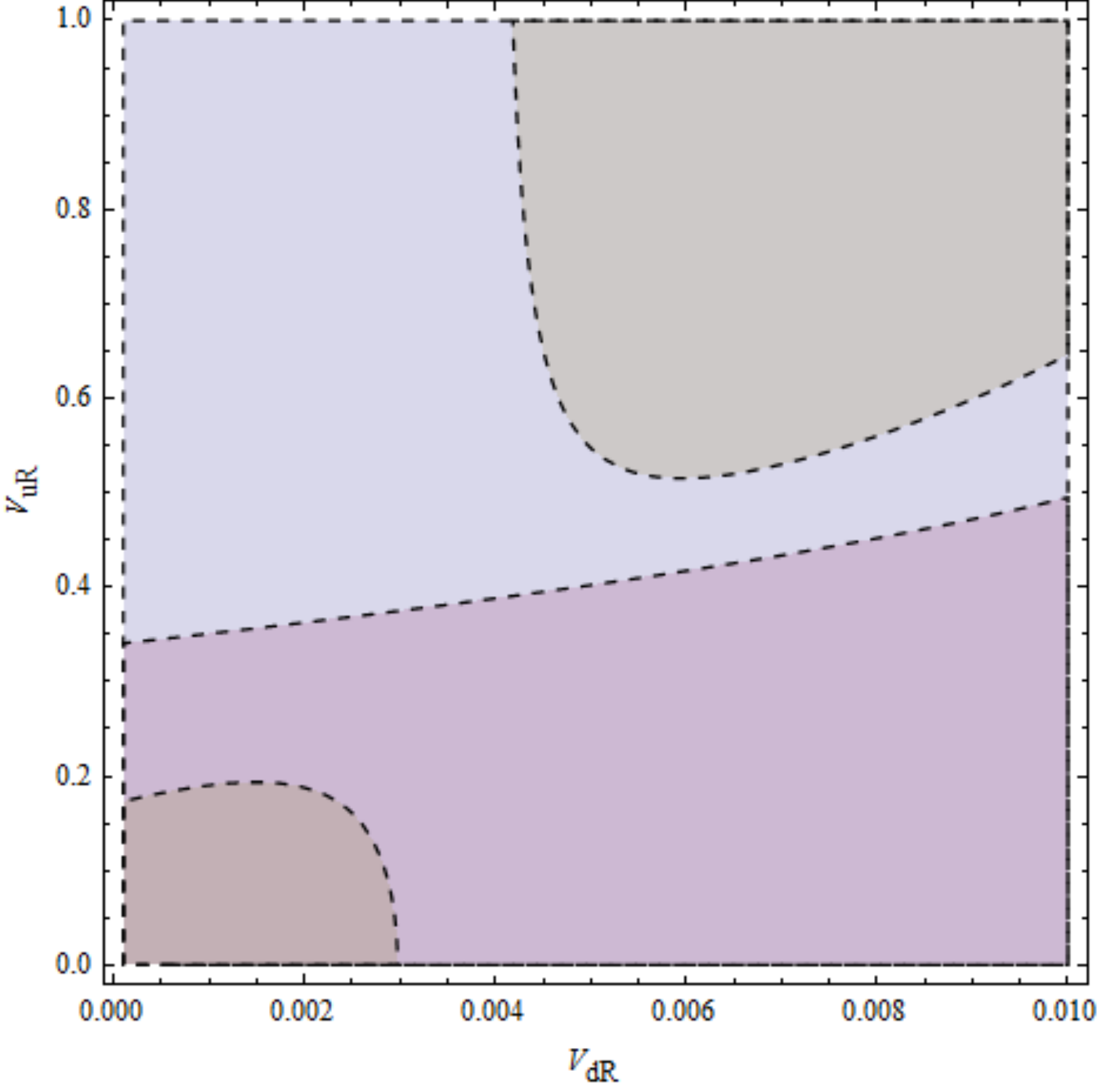} 
  \end{tabular}
   \caption{\label{vuvd} The left panel presents constraints for ($V_{uR},V_{dR}$) coming from the meson mass differences, $\Delta m_{K,B_d,B_s}$,  with respect to the new physics scale, $M=5$ TeV, while the right panel is those for the new physics scale, $M=10$ TeV.}
\end{figure}

In Fig. \ref{vuvd}, we make contours for the mass differences, $\Delta m_K$, $\Delta m_{B_d}$, and $\Delta m_{B_s}$, as functions of the right-handed quark mixing matrix elements $(V_{uR},V_{dR})$ for the new physics scale $M=5$ TeV and $M=10$ TeV, respectively. The $M$ values have been chosen consistently with the bounds previously given. The available region for $\Delta m_K$ is the whole frame. The two separated regions are for $\Delta m_{B_d}$. A lower half region is for $\Delta m_{B_s}$. Hence, the available parameter space for $\Delta m_{K,B_d,B_s}$ is only the region (darkest) at the left-down corner for each panel. From the allowed regimes, we obtain constraints for the right-handed quark
mixing matrix elements as $|V_{uR}|<0.08$ and $|V_{dR}|<0.0015$ for $M=5$ TeV, while $|V_{uR}|<0.2$ and $|V_{dR}|<0.003$ are for $M=10$ TeV.

\begin{figure}[H]
 \centering
 \begin{tabular}{ccc}
\includegraphics[width=5cm]{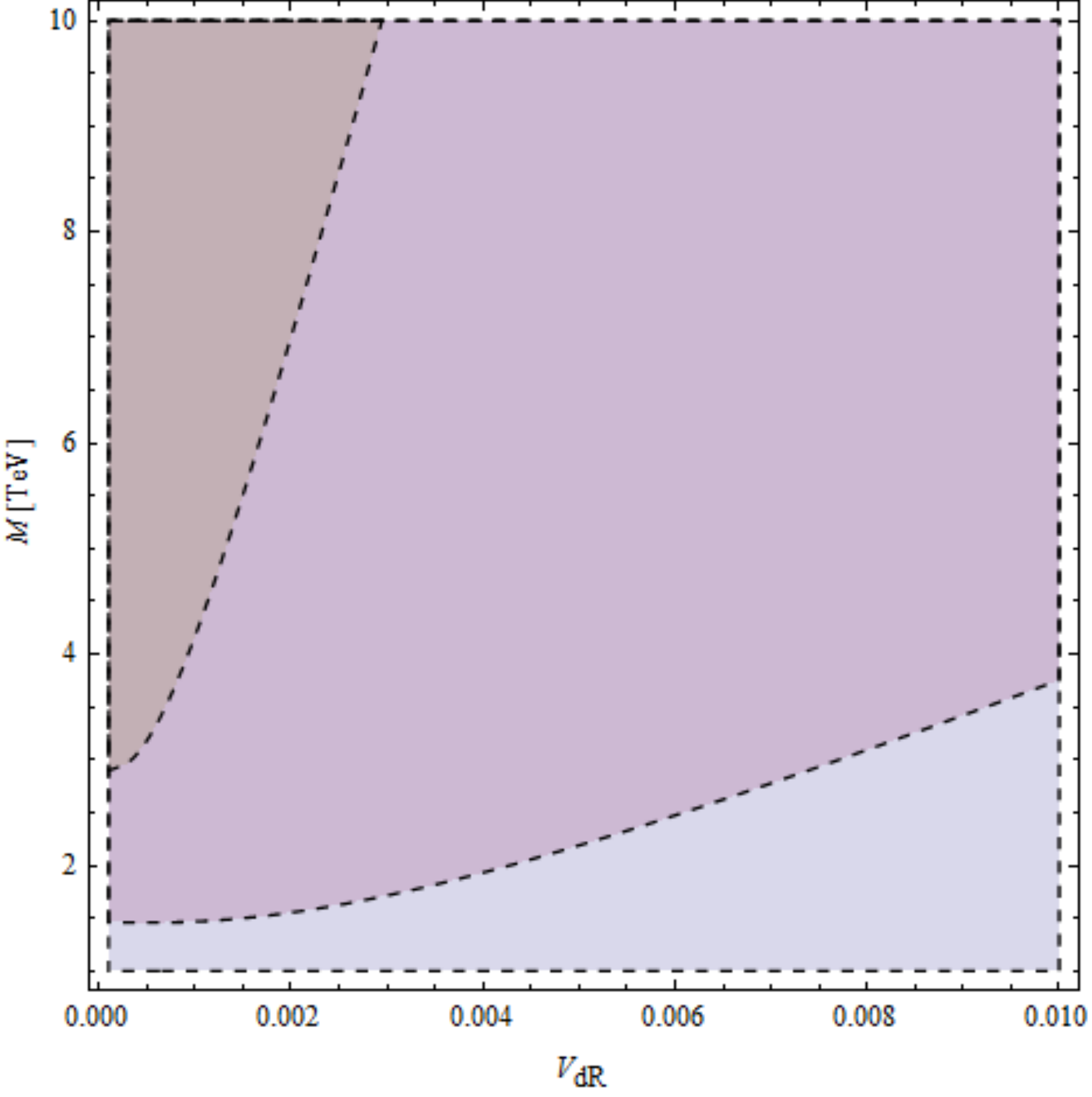} &
\includegraphics[width=5cm]{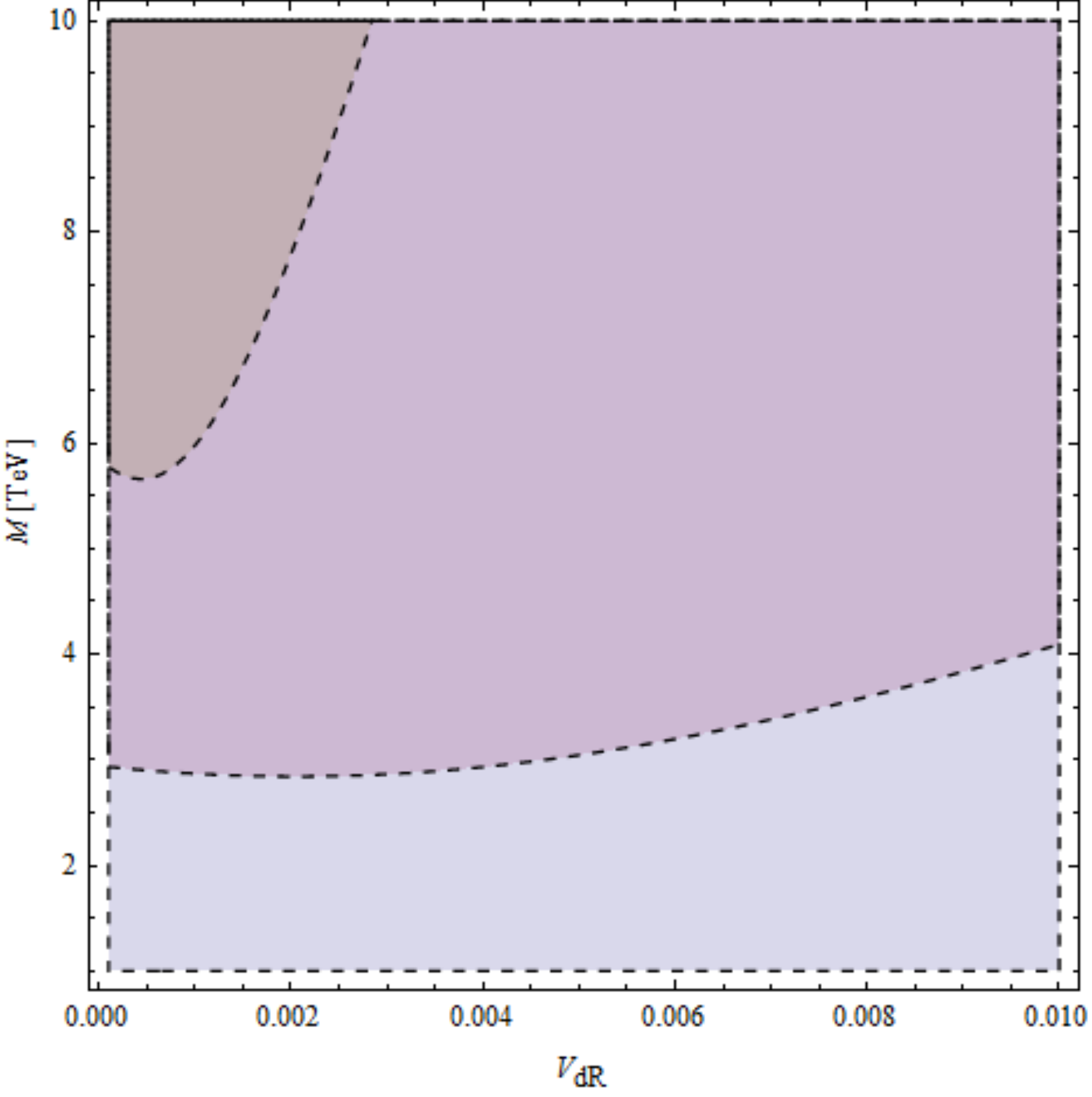} &
\includegraphics[width=5cm]{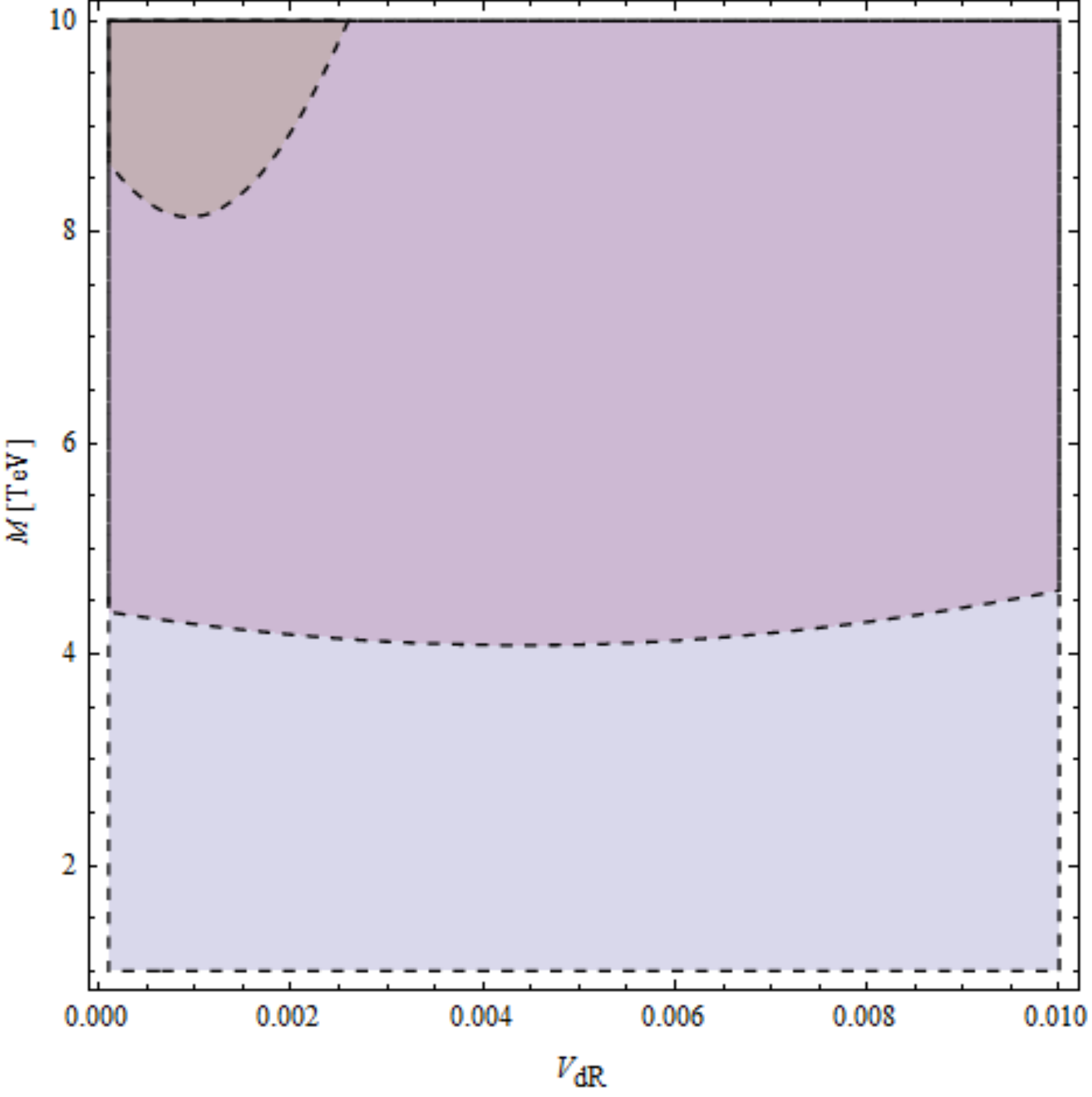}  
\end{tabular}
\caption{\label{dmkds} The left, middle, right panels present bounds for ($M,V_{dR}$) coming from the meson mass differences, $\Delta m_{K,B_d,B_s}$, corresponding to $V_{uR}=0.05,\ 0.1$, and 0.15, respectively.}
\end{figure}

Considering $V_{uR}=0.05,\ 0.1$, and 0.15, we make contours for $\Delta m_K$, $\Delta m_{B_d}$, and $\Delta m_{B_s}$ as functions of $(M, V_{dR})$ in Fig.
\ref{dmkds}, respectively. The viable parameter space is the region (darkest) bounded at left-upper corner for each panel. We obtain $M>2.8$ TeV for $V_{uR}=0.05$ (left panel), $M>5.7$ TeV for $V_{uR}=0.1$ (middle panel), and $M>8.2$ TeV for $V_{uR}=0.15$ (right panel). Thus the new physics scale $M$ is low when $V_{uR}$ is low, and vice versa.

We see that the bounds for the $H_2$ and $Z'_R$ masses consistently with the new physics scale given in the previous subsection. 

\section{\label{conclusion}Conclusion}

We have shown that the left-right asymmetric model with $SU(3)_C\otimes SU(2)_L\otimes SU(3)_R\otimes U(1)_X$ gauge group naturally provides the new, tree-level FCNCs through both gauge and Yukawa interactions as a result of the non-universal fermion generations, which is different from the minimal left-right symmetric model. The new gauge symmetry contains automatically not only the right-handed neutrinos but also
the wrong $B-L$ particles which induce the observed neutrino masses and dark matter candidates as a result of the gauge symmetry breaking. Particularly, the W-parity which is actually larger than $Z_2$ and stabilizing the dark matter is naturally realized as a residual gauge symmetry. In other words, they all arise from the gauge principles.   

The scalar sector has been explicitly diagonalized. The number of the Goldstone bosons matches the number of the massive
gauge bosons. There are twelve physical scalar fields, one of which is the standard model Higgs boson and the others are new and heavy. Because of the condition, $u,v \ll w, \Lambda$, the standard model Higgs boson gains a mass at the leading order in the electroweak scale, and it slightly mixes with the new neutral Higgs bosons. The gauge sector has been explicitly diagonalized too. The model contains five new heavy gauge bosons, $\mathcal{Z}_1,\ \mathcal{Z}'_1,\ W^\pm_2,\ X_R^{\pm q},\ Y^{\pm (q+1)}_R$, besides the standard model like gauge bosons, $A,\ \mathcal{Z},\ W^\pm_1$. The charged
gauge bosons, $W^\pm_{L,R}$, mix via a small angle, $t_{\xi} \propto \frac{uv}{\Lambda^2}$. Also, the neutral gauge boson $Z_L$ slightly mix with the new neutral gauge bosons, $Z_R$ and $Z'_R$, which is suppressed by $u,v\ll w,\La$ too. In the $B_{d,s}$ and $K$ mass differences, the model can have box diagrams due to the contributions of $W^\pm_2$, $H^\pm_5$, and even other new particles, in addition to the standard model boxes, but such contributions are more suppressed by $u,v\ll w,\La,f$. Furthermore, the new FCNCs that come from the tree level interactions with $Z'_R$ and $H_2$ are larger than the mentioned ones by loop factors.       

All the interactions of the gauge bosons with fermions and scalars
have been derived. The standard model interactions are successfully recovered.
The new interactions play important rules, which change quark flavors as well as set dark matter observables, besides others. We
have concentrated on the first kind interactions as induced by $Z'_R$ and $H_2$, and obtaining their contributions to the neutral meson mass differences, $\Delta m_{K,B_d,B_s}$, which depend on the new particle masses and the elements of the right-handed quark mixing matrices. The mixing effects also modify $\rho$-parameter as well as the well-measured couplings of $W,Z$ bosons which are determined by the mixing parameters $\xi,\epsilon_1,\epsilon_2$. In agreement with electroweak precision measurements, the parameters $\rho$, $\xi$, and $\epsilon_{1,2}$ set the new physics scale (assuming $w=\La$) as $4.6\ \mathrm{TeV}<\La<13.7\ \mathrm{TeV}, 5.5\ \mathrm{TeV}<\La< 16.3\ \mathrm{TeV}$, and $6.6\ \mathrm{TeV}<\La<19.4\ \mathrm{TeV}$ for $\beta=1/\sqrt{3},\ 0$, and $-1/\sqrt{3}$, respectively. It also set the narrow regimes for the weak scales such as $u>222.3$, $215$, and $210.4\ \mathrm{GeV}$ for those respective $\beta$ values [the $v$ scale is thus followed from $u^2+v^2=(246\ \mathrm{GeV})^2$, and noting that $u,v<246$ GeV]. The mass differences yield that when the new physics masses are fixed, the right-handed quark mixing elements are constrained, such that $|V_{uR}|<0.08$ and $|V_{dR}|<0.0015$ for $M=5$ TeV, while $|V_{uR}|<0.2$ and $|V_{dR}|<0.003$ for $M=10$ TeV, assuming that $m_{Z'_R}=m_{H_2}\equiv M$ and $(V_{dR})_{31}=(V_{dR})_{32}\equiv V_{dR}$ and $(V_{uR})_{33}\equiv V_{uR}$ for short. In other case, fixing $V_{uR}=0.05$, 0.1, and 0.15, we obtain $M>2.8$ TeV, $M>5.7$ TeV, and $M>8.2$ TeV, respectively, where $V_{dR}$ is free to float. It yields that the new physics scale is more sensitive to $V_{uR}$. The conclusion is that the two kinds of bounds are compatible, and the new physics scale should be in 5--10 TeV order.

This model may predict the quantization of charges such as the electric charge and $B-L$. It belongs to a class of the model that provides dark matter naturally without supersymmetry. All these are worth exploring to be published elsewhere \cite{dhip}.

\section*{Acknowledgements}

This research is funded by Vietnam National Foundation for Science and Technology Development (NAFOSTED) under grant number 103.01-2016.77.  

\appendix 

\section{\label{gvaf} Vector and axial-vector couplings}

This section obtains all the couplings of fermions with the neutral gauge bosons $Z$, $\mathcal{Z}_1$, and $\mathcal{Z}'_1$ as displayed in Tables \ref{ttZ}, \ref{ttZ1}, and \ref{ttZ1p}, respectively.  

\begin{table}[!h]
\bc
\begin{tabular}{|c|c|c|c|c|c|}
\hline
$f$ & $g^{Z}_V(f)$ & $g^{Z}_A(f)$&$f$ & $g^{Z}_V(f)$ & $g^{Z}_A(f)$ \\ 
\hline
$\nu_a$ &  $\fr 1 2$ & $\fr 1 2$ & $e_a$ & $-\fr 1 2 + 2 s^2_W$ & $-\fr 1 2 $ \\
\hline
$E_a$ & $-2s^2_Wq$ &  $0$ &$u_a$ & $\fr 1 2 -\fr 4 3 s^2_W$ & $\fr 1 2$ \\
\hline
$d_a$ & $-\fr 1 2 + \fr 2 3 s^2_W $ & $-\fr 1 2$ &$J_{\alpha}$ & $2s^2_W (q+\fr 1 3)  $ & $0$ \\
\hline
$J_{3}$ & $-2s^2_W (q+\fr 2 3)  $ & $0$ & & &\\
\hline
\end{tabular}
\caption{\label{ttZ}The couplings of $Z$ with fermions.}  
\ec
\end{table}

\begin{table}[!h]
\bc
\begin{tabular}{|c|c|c|}
\hline
$f$ & $g^{\mathcal{Z}_{1}}_V(f)$ & $g^{\mathcal{Z}_{1}}_A(f)$ \\
\hline 
$\nu_a $ & $\fr{t_X[t_R^2+\beta t_X^2(2\sqrt{3}+\beta)]c_\epsilon c_W+\sqrt{3}[t_R^2+t_X^2(2+\beta^2)]s_\epsilon s_W}{2\sqrt{3} t_X\sqrt{t_R^2+t_X^2\beta^2}}$ &  $-\fr{\sqrt{t_R^2+t_X^2\beta^2}(t_Xc_\epsilon c_W+\sqrt{3}s_\epsilon s_W)}{2\sqrt{3}t_X}$\\
\hline
$e_a$ & $\fr{t_X[t_R^2+\beta t_X^2(2\sqrt{3}+\beta)]c_\epsilon c_W-\sqrt{3}[t_R^2+t_X^2(\beta^2-2)]s_\epsilon s_W}{2\sqrt{3} t_X\sqrt{t_R^2+t_X^2\beta^2}}$ &$-\fr{\sqrt{t_R^2+t_X^2\beta^2}(t_Xc_\epsilon c_W-\sqrt{3}s_\epsilon s_W)}{2\sqrt{3}t_X}$\\
\hline 
$E_a$ & $\fr{-(t_R^2+\beta^2 t_X^2)c_\epsilon c_W-2\sqrt{3}qt_X(t_X\beta c_\epsilon c_W+s_\epsilon s_W)}{\sqrt{3} \sqrt{t_R^2+t_X^2\beta^2}}$& $\fr{\sqrt{t_R^2+t_X^2\beta^2}c_\epsilon c_W}{\sqrt{3}}$\\
\hline 
$u_\al$ &$\fr{-t_X[\sqrt{3}t_R^2+\beta t_X^2(2+\sqrt{3}\beta)]c_\epsilon c_W+[3t_R^2+t_X^2(3\beta^2-2)]s_\epsilon s_W}{6 t_X\sqrt{t_R^2+t_X^2\beta^2}}$&$\fr{\sqrt{t_R^2+t_X^2\beta^2}(t_Xc_\epsilon c_W-\sqrt{3}s_\epsilon s_W)}{2\sqrt{3}t_X}$\\
\hline 
$u_3$  &$\fr{t_X[\sqrt{3}t_R^2+\beta t_X^2(\sqrt{3}\beta-2)]c_\epsilon c_W+[3t_R^2+t_X^2(3\beta^2-2)]s_\epsilon s_W}{6 t_X\sqrt{t_R^2+t_X^2\beta^2}}$ & $-\fr{\sqrt{t_R^2+t_X^2\beta^2}(t_Xc_\epsilon c_W+\sqrt{3}s_\epsilon s_W)}{2\sqrt{3}t_X}$ \\
\hline 
$d_\al$ &$\fr{-t_X[\sqrt{3}t_R^2+\beta t_X^2(2+\sqrt{3}\beta)]c_\epsilon c_W-[3t_R^2+t_X^2(3\beta^2+2)]s_\epsilon s_W}{6 t_X\sqrt{t_R^2+t_X^2\beta^2}}$ & $\fr{\sqrt{t_R^2+t_X^2\beta^2}(t_Xc_\epsilon c_W+\sqrt{3}s_\epsilon s_W)}{2\sqrt{3}t_X}$\\
\hline
$d_3$ & $\fr{t_X[\sqrt{3}t_R^2+\beta t_X^2(\sqrt{3}\beta-2)]c_\epsilon c_W-[3t_R^2+t_X^2(3\beta^2+2)]s_\epsilon s_W}{6 t_X\sqrt{t_R^2+t_X^2\beta^2}}$ &  $-\fr{\sqrt{t_R^2+t_X^2\beta^2}(t_Xc_\epsilon c_W-\sqrt{3}s_\epsilon s_W)}{2\sqrt{3}t_X}$ \\
\hline 
$J_\al$ & $\fr{[\sqrt{3}t_R^2+\beta t_X^2(2+6q+\sqrt{3}\beta)]c_\epsilon c_W+2(1+3q)t_X s_\epsilon s_W}{3\sqrt{t_R^2+t_X^2\beta^2}}$ & $-\fr{\sqrt{t_R^2+t_X^2\beta^2}c_\epsilon c_W}{\sqrt{3}}$ \\
\hline
$J_3$ & $\fr{-\sqrt{3}(t_R^2+\beta^2 t_X^2)c_\epsilon c_W-2(2+3q)t_X(t_X\beta c_\epsilon c_W+s_\epsilon s_W)}{3\sqrt{t_R^2+t_X^2\beta^2}}$& $\fr{\sqrt{t_R^2+t_X^2\beta^2}c_\epsilon c_W}{\sqrt{3}}$  \\
\hline
\end{tabular}
\caption{\label{ttZ1} The couplings of $\mathcal{Z}_{1}$ with fermions}
\ec
\end{table}  

\begin{table}[!h]
\bc
\begin{tabular}{|c|c|c|}
\hline
$f$ & $g^{\mathcal{Z}_{1}^\prime}_V(f)$ & $g^{\mathcal{Z}_{1}^\prime}_A(f)$ \\
\hline 
$\nu_a $ & $\fr{t_X[t_R^2+\beta t_X^2(2\sqrt{3}+\beta)]s_\epsilon c_W-\sqrt{3}[t_R^2+t_X^2(2+\beta^2)]c_\epsilon s_W}{2\sqrt{3} t_X\sqrt{t_R^2+t_X^2\beta^2}}$ &  $-\fr{\sqrt{t_R^2+t_X^2\beta^2}(t_Xs_\epsilon c_W-\sqrt{3}c_\epsilon s_W)}{2\sqrt{3}t_X}$\\
\hline
$e_a$ & $\fr{t_X[t_R^2+\beta t_X^2(2\sqrt{3}+\beta)]s_\epsilon c_W+\sqrt{3}[t_R^2+t_X^2(\beta^2-2)]c_\epsilon s_W}{2\sqrt{3} t_X\sqrt{t_R^2+t_X^2\beta^2}}$ &$-\fr{\sqrt{t_R^2+t_X^2\beta^2}(t_Xs_\epsilon c_W+\sqrt{3}c_\epsilon s_W)}{2\sqrt{3}t_X}$\\
\hline 
$E_a$ & $\fr{-(t_R^2+\beta^2 t_X^2)s_\epsilon c_W-2\sqrt{3}qt_X(t_X\beta s_\epsilon c_W-c_\epsilon s_W)}{\sqrt{3} \sqrt{t_R^2+t_X^2\beta^2}}$& $\fr{\sqrt{t_R^2+t_X^2\beta^2}s_\epsilon c_W}{\sqrt{3}}$\\
\hline 
$u_\al$ &$\fr{-t_X[\sqrt{3}t_R^2+\beta t_X^2(2+\sqrt{3}\beta)]s_\epsilon c_W-[3t_R^2+t_X^2(3\beta^2-2)]c_\epsilon s_W}{6 t_X\sqrt{t_R^2+t_X^2\beta^2}}$&$\fr{\sqrt{t_R^2+t_X^2\beta^2}(t_Xs_\epsilon c_W+\sqrt{3}c_\epsilon s_W)}{2\sqrt{3}t_X}$\\
\hline 
$u_3$  &$\fr{t_X[\sqrt{3}t_R^2+\beta t_X^2(\sqrt{3}\beta-2)]s_\epsilon c_W-[3t_R^2+t_X^2(3\beta^2-2)]c_\epsilon s_W}{6 t_X\sqrt{t_R^2+t_X^2\beta^2}}$ & $-\fr{\sqrt{t_R^2+t_X^2\beta^2}(t_Xs_\epsilon c_W-\sqrt{3}c_\epsilon s_W)}{2\sqrt{3}t_X}$ \\
\hline 
$d_\al$ &$\fr{-t_X[\sqrt{3}t_R^2+\beta t_X^2(2+\sqrt{3}\beta)]s_\epsilon c_W+[3t_R^2+t_X^2(3\beta^2+2)]c_\epsilon s_W}{6 t_X\sqrt{t_R^2+t_X^2\beta^2}}$ & $\fr{\sqrt{t_R^2+t_X^2\beta^2}(t_Xs_\epsilon c_W-\sqrt{3}c_\epsilon s_W)}{2\sqrt{3}t_X}$\\
\hline
$d_3$ & $\fr{t_X[\sqrt{3}t_R^2+\beta t_X^2(\sqrt{3}\beta-2)]s_\epsilon c_W+[3t_R^2+t_X^2(3\beta^2+2)]c_\epsilon s_W}{6 t_X\sqrt{t_R^2+t_X^2\beta^2}}$ &  $-\fr{\sqrt{t_R^2+t_X^2\beta^2}(t_Xs_\epsilon c_W+\sqrt{3}c_\epsilon s_W)}{2\sqrt{3}t_X}$ \\
\hline 
$J_\al$ & $\fr{[\sqrt{3}t_R^2+\beta t_X^2(2+6q+\sqrt{3}\beta)]s_\epsilon c_W-2(1+3q)t_X c_\epsilon s_W}{3\sqrt{t_R^2+t_X^2\beta^2}}$ & $-\fr{\sqrt{t_R^2+t_X^2\beta^2}s_\epsilon c_W}{\sqrt{3}}$ \\
\hline
$J_3$ & $\fr{-\sqrt{3}(t_R^2+\beta^2 t_X^2)s_\epsilon c_W-2(2+3q)t_X(t_X\beta s_\epsilon c_W-c_\epsilon s_W)}{3\sqrt{t_R^2+t_X^2\beta^2}}$& $\fr{\sqrt{t_R^2+t_X^2\beta^2}s_\epsilon c_W}{\sqrt{3}}$  \\
\hline
\end{tabular}
\caption{\label{ttZ1p} The couplings of $\mathcal{Z}_{1}^\prime$ with fermions}
\ec
\end{table}


\begin{thebibliography}{99}

\bibitem {data} K. A. Olive {\it et al.} (Particle Data Group), Chin. Phys. C {\bf 38}, 090001 (2014), and partial updates at http://pdg.lbl.gov.

\bibitem{bsmumu} V. Khachatryan {\it et al.} (CMS and LHCb Collaborations), Nature {\bf 522}, 68 (2015).

\bibitem{lhcb1} R. Aaij {\it et al.} (LHCb Collaboration), JHEP {\bf 02}, 104 (2016). 

\bibitem{lhcb2} R. Aaij {\it et al.} (LHCb Collaboration), JHEP {\bf 09}, 179 (2015). 

\bibitem{lhcb3} R. Aaij {\it et al.} (LHCb Collaboration), Phys. Rev. Lett. {\bf 113}, 151601 (2014). 

\bibitem{lhcbcombi} S. Descotes-Genon {\it et al.}, JHEP {\bf 06}, 092 (2016). 

\bibitem{LR1} J. C. Pati and A. Salam, Phys. Rev. D \textbf{10}, 275 (1974); R. N. Mohapatra and J. C. Pati,
Phys. Rev. D \textbf{11}, 566 (1975); R. N. Mohapatra and J. C. Pati, Phys. Rev. D \textbf{11}, 2558
(1975); G. Senjanovi\'c and R. N. Mohapatra, Phys. Rev. D \textbf{12}, 1502 (1975); G. Senjanovi\'c, Nucl.
Phys. B \textbf{153}, 334 (1979).

\bibitem{NLR} P. Minkowski, Phys. Lett. B \textbf{67}, 421 (1977);
 R. N. Mohapatra and G. Senjanovi\'c, Phys. Rev. Lett. \textbf{44}, 912 (1980);  R. N. Mohapatra and G. Senjanovi\'c,
Phys. Rev. D \textbf{23}, 165 (1981).

\bibitem{PheLR} G. Beall, M. Bander, and A. Soni, Phys. Rev. Lett. \textbf{48}, 848
(1982), R. N. Mohapatra, G. Senjanovich, and M. Tran, Phys. Rev. D\textbf{28}, 546 (1983); G. Ecker,
W. Grimus, and H. Neufeld, Phys. Lett. B \textbf{127}, 365 (1983);  F. G. Gilman and M. H.
Reno, Phys. Lett. B \textbf{127}, 426 (1983); F. G. Gilman and M. H. Reno, Phys. Rev. D \textbf{29},
937 (1983), G. Ecker and W. Grimus, Nucl. Phys. B \textbf{258}, 328 (1985); J. -M. Frere {\it et al.}, Phys. Rev. D {\bf 46}, 337 (1992); 
M. E. Pospelov, Phys. Rev. D \textbf{56}, 259 (1997) [arXiv:hep-ph/9611422]; A. Maiezza {\it et al.}, Phys. Rev. D {\bf 82}, 055022 (2010). 

\bibitem{vubpro} A. Crivellin, Phys. Rev. D {\bf 81}, 031301 (2010); A. J. Buras, K. Gemmler, and G. Isidori, Nucl. Phys. B {\bf 843}, 107 (2011); M. Blanke, A. J. Buras, K. Gemmler, and T. Heidsieck, JHEP {\bf 03}, 024 (2012). 

\bibitem{exdiphoton} The ATLAS Collaboration, ATLAS-CONF-2015-081; The CMS Collaboration, CMS-PAS-EXO-15-004.

\bibitem{noexdiphoton} The ATLAS Collaboration, ATLAS-CONF-2016-059; The CMS Collaboration, CMS-PAS-EXO-16-027. 

\bibitem{diphoton} F. F. Deppisch, C. Hati, S. Patra, P. Pritimita, and U. Sarkar,
Phys. Lett. B \textbf{757}, 223 (2016) [arXiv:1601.00952 [hep-ph]]; C. Hati, Phys. Rev. D \textbf{93}, 075002 (2016) [arXiv:1601.02457 [hep-ph]; 
 C.-H. Chen and T. Nomura, arXiv:1606.03804 [hep-ph]. 

\bibitem{3L3R} D. T. Huong and P. V. Dong, Phys. Rev. D {\bf 93}, 095019 (2016) [arXiv:1603.05146 [hep-ph]].

\bibitem{d3311} P. V. Dong, T. D. Tham, and H. T. Hung, Phys. Rev. D {\bf 87}, 115003 (2013); P. V. Dong, D. T. Huong, F. S. Queiroz, and N. T. Thuy, Phys. Rev. D {\bf 90}, 075021 (2014); D. T. Huong, P. V. Dong, C. S. Kim, and N. T. Thuy, Phys. Rev. D {\bf 91}, 055023 (2015); P. V. Dong, Phys. Rev. D {\bf 92}, 055026 (2015); P. V. Dong and D. T. Si, Phys. Rev. D {\bf 93}, 115003 (2016); D. T. Huong and P. V. Dong, arXiv:1605.01216 [hep-ph]. 

\bibitem{331} F. Pisano and V. Pleitez, Phys. Rev. D {\bf 46}, 410 (1992); P. H. Frampton, Phys. Rev. Lett. {\bf 69}, 2889
(1992); R. Foot, O. F. Hernandez, F. Pisano, and V. Pleitez, Phys. Rev. D {\bf 47}, 4158 (1993),
M. Singer, J. W. F. Valle, and J. Schechter, Phys. Rev. D {\bf 22}, 738 (1980); J. C. Montero, F. Pisano,
and V. Pleitez, Phys. Rev. D {\bf 47}, 2918 (1993); R. Foot, H. N. Long, and Tuan A. Tran, Phys. Rev. D
{\bf 50}, 34 (1994); P. V. Dong, H. N. Long, D. T. Nhung, and D. V. Soa, Phys. Rev. D {\bf 73}, 035004 (2006);
S. M. Boucenna, J. W. F. Valle, and A. Vicente, Phys. Rev. D {\bf 92}, 053001 (2015); J. W. F. Valle and
C. A. Vaquera-Araujo, Phys. Lett. B {\bf 755}, 363 (2016).

\bibitem{dhip} P. V. Dong and D. T. Huong, ``Left-right asymmetry and dark matter'', work in progress.     

\bibitem{chargequan}P. V. Dong and H. N. Long, Eur. Phys. J. C {\bf 42}, 325
(2005) [arXiv:hep-ph/0604028].

\bibitem{Higgsd} G. Aad {\it et al.} (ATLAS Collaboration), Phys. Lett. B {\bf 716}, 1 (2012); S. Chatrchyan {\it et al.} (CMS Collaboration), Phys. Lett. B {\bf 716},30 (2012); G. Aad {\it et al.} (ATLAS Collaboration), arXiv:1507.04548; S. Chatrchyan {\it et al.} (CMS Collaboration), CMS-PAS-HIG-14-009. 

\bibitem{matrixelements} F. Gabbiani, E. Gabrielli, A. Masiero, and L. Silvestrini, Nucl. Phys. B {\bf 477}, 321 (1996)
[arXiv:hep-ph/9604387].

\bibitem{data1} Y. Amhis {\it et al.} (Heavy Flavor Averaging Group (HFAG) Collaboration), arXiv:1412.7515 [hep-ex].

\bibitem{paras} A. J. Buras {\it et al.}, JHEP {\bf 06}, 111 (2013). 

\bibitem{dmkbdsm} A. Lenz, arXiv:1603.07770 [hep-ph]; A. J. Buras and F. D. Fazio, arXiv:1604.02344 [hep-ph].  

\bibitem{close331} A. G. Dias, J. C. Montero, and V. Pleitez, Phys. Lett. B {\bf 637}, 85 (2006); Phys. Rev. D {\bf 73}, 113004 (2006).  

\bibitem{landau} A. G. Dias, R. Martinez, and V. Pleitez, Eur. Phys. J. C {\bf 39}, 101 (2005); A. G. Dias, Phys. Rev. D {\bf 71}, 015009 (2005).   

\end{thebibliography}
 \end{document}